\documentclass[12pt]{iopart}
\usepackage{iopams}
\usepackage{graphicx}
\begin{document}

\title[ Model C critical dynamics of disordered magnets]{ Model C critical dynamics of disordered magnets}

\author{Maxym Dudka$^{1,2}$
, Reinhard Folk$^2$, Yurij Holovatch$^{1,2,3}$ and G\"unter
Moser$^{4}$
 }

\address{$^1$ Institute for Condensed Matter Physics, National
Academy of Sciences of Ukraine, UA--79011 Lviv, Ukraine}

\address{$^2$ Institut f\"ur Theoretische Physik, Johannes
Kepler Universit\"at Linz, A-4040, Linz, Austria}

\address{$^3$ Ivan Franko National University of Lviv, UA--79005
Lviv, Ukraine}

\address{$^4$ Institut f\"ur Physik und Biophysik, Univerit\"at
Salzburg, A--5020 Salzburg, Austria}

\eads{\mailto{maxdudka@icmp.lviv.ua},
\mailto{folk@tphys.uni-linz.ac.at}, \mailto{hol@icmp.lviv.ua}}

\begin{abstract}
The critical dynamics of model C in the presence of disorder is
considered. It is known that in the asymptotics a conserved
secondary density decouples from  the nonconserved order parameter
for disordered systems. However couplings between order parameter
and secondary density cause considerable effects on non-asymptotic
critical properties. Here,  a general procedure for a
renormalization group treatment is proposed. Already  the one-loop
approximation  gives a qualitatively correct picture of the
diluted model C dynamical criticality. A more quantitative
description is achieved using two-loop approximation. In order to
get reliable results resummation technique has to be applied.
\end{abstract}


\submitto{\JPA}

\section{Introduction}
The relevance of weak structural disorder for the critical
behaviour of pure magnetic systems has been in the focus of
attention of researchers since a long time \cite{review}.
Depending on the structure of  a magnet disorder can enter as
random fields, random anisotropies, random bond or site dilution.
Our special interest here is in diluted systems. Concerning the
statics of such systems one of the main insights comes from the
Harris criterion \cite{Harris74} stating that the static
asymptotic critical behaviour remains unchanged, if the specific
heat of the pure system does not diverge \cite{note1}. Since the
borderline value $n_c$ between a diverging and nondiverging
specific heat in space dimension $d=3$ lies between order
parameter (OP) dimension $n=1$ (Ising model) and $n=2$ (XY model)
\cite{Bervillier86} only
 diluted Ising systems show a new critical behaviour, which is
characterized by a non-diverging specific heat \cite{review}.

Systems with the same static critical behaviour may have different
dynamic critical properties. In the vicinity of a critical point
slow processes play an important role, therefore in  critical
dynamics the main objects are the slow modes. Apart from the order
parameter these are the conserved quantities, which  coupled to
the OP. The simplest dynamical model - model C \cite{Halperin74} -
includes these couplings only statically. For such a case the
Harris criterion also has consequences. It leads to the conclusion
that the coupling of conserved quantities to the OP in disordered
systems is of no relevance \cite{Krey76,Krey77,Lawrie84}. The
argumentation of this statement is based on the fact
\cite{Halperin74,FoMo03,Folk04} that coupling to
 a conserved density for the relaxational model is relevant
only if the specific heat diverges. Diluted models always have
nondiverging specific heat and therefore  their coupling beetwen a
conserved density and OP is irrelevant. As a consequence most of
the studies of critical dynamics considered only relaxational
dynamics of diluted Ising systems
\cite{Grinstein77,Prudnikov92,Oerding95,Janssen95}.

However the conclusion concerning the irrelevance of the conserved
density is made  only for the asymptotic properties of the model.
Meanwhile it became clear that in diluted magnets most of the
experiments and simulations are in the nonasymptotic region
\cite{review,Dudka03,Perumal03,Berche04,Calabrese04,Berche05}.
Within the standard tool of investigation of critical phenomena,
namely the field-theoretical renormalization group (RG), the
nonasymptotic critical behaviour is described by the flow of the
model parameters  (static and dynamic) leading to effective
critical behaviour rather than to the asymptotic one characterized
by power laws with universal exponents determined by the fixed
point properties of the parameters.

Thus we consider the dynamics of a diluted magnetic system where
the OP relaxes with a relaxation rate $\Gamma$ and the second
conserved density shows diffusive behaviour with a diffusion rate
$\lambda$. The ratio of the time scales $w=\Gamma/\lambda$ is the
dynamic parameter of  model C and takes on  different fixed point
values depending on the dimension $n$ of the OP  and spatial
dimension $d$.  Only recently the correct field theoretic RG
functions of model C have been calculated \cite{FoMo03,Folk04}. It
turned out that model C itself has a very slow transient at least
for the cases $n=1,2$ in two-loop order.

Another artifact present in the RG analysis of diluted magnets
within one loop order is the  degeneracy of the static
$\beta$-functions for $n=1$. As it is well established now it
leads to a $\sqrt{4-d}$ rather than a $4-d$-expansion
\cite{sqrteps,Grinstein78}. This makes a two-loop treatment
inevitable. Moreover the borderline where the specific heat
exponent $\alpha$ changes in two-loop order is shifted from $n=4$
to below $n=2$.

Thus two-loop calculations are necessary (i) to specify the shift
of the stability regions and (ii) to calculate quantitative values
of the effective exponents.

 All the above arguments
serve as a reason to consider the disordered model C critical
dynamics by the field-theoretical RG approach within two-loop
approximation. Being more elaborated technically this
approximation should lead to reliable numerical results which as
previous experience shows should not be changed essentially by
further increase of the perturbation theory order. Some of our
results are briefly summarized in Ref. \cite{latter}, and here we
give  a more thorough derivation of the dynamic RG function and
analyze the diluted model C effective critical dynamics.

The set-up of the paper is the following: in the next Section
 we define the dynamical model for disordered magnets, its
renormalization and general relations for the field theoretic
functions. Results in one-loop order are collected in  Section
\ref{III}. Two-loop results are presented in Section \ref{IV}.
Conclusions are given in Section \ref{V}.

\section{\label{II} Model and renormalization}

\subsection{Model equations}
The object of our analysis consists in a dynamical model for
quenched { disordered} magnets, namely model C in the
classification of Ref.\cite{Halperin77}.  Model C describes the
dynamics of a nonconserved OP $\vec{\varphi}_0$ which is coupled
to a conserved density $m_0$ (in most cases the energy density).
The secondary density has to be taken into account since it is
also a slow density  showing critical slowing down near the phase
transition. The OP is assumed to be an $n$-component vector, while
the density $m_0$ is a scalar quantity.  The structure of the
equations \cite{Halperin77,Halperin74} of motion is not changed by
the presence of disorder. They read:
\begin{eqnarray}\label{eq_mov2}
\frac{\partial {\varphi}_{i,0}}{\partial
t}&=&-\mathring{\Gamma}\frac{\partial {\mathcal H}}{\partial
{\varphi}_{i,0}}+{\theta}_{{\varphi}_i}\qquad i=1\ldots n, \\
\label{eq_mov2v}
 \frac{\partial {m}_0}{\partial
t}&=&\mathring{\lambda}_m\nabla^2\frac{\partial {\mathcal
H}}{\partial {m}_0}+{\theta_{{m}}}.
\end{eqnarray}

The OP relaxes to equilibrium with the relaxation rate
$\mathring{\Gamma}$ and the conserved density $m_0$ diffuses with
the diffusion rate $\mathring{\lambda}_m$. The stochastic forces
in (\ref{eq_mov2}), (\ref{eq_mov2v}) satisfy the Einstein
relations:
\begin{eqnarray}\label{1}
<{\theta}_{\varphi_i}(x,t){\theta}_{\varphi_j}(x',t')>&=&2\mathring{\Gamma}\delta(x-x')\delta(t-t')\delta_{ij},
\\ \label{2}
 <{\theta}_{m}(x,t){\theta}_{{m}}(x',t')>&=&-2\mathring{\lambda}_{{m}}\nabla^2\delta(x-x')\delta(t-t')\delta_{ij}.
\end{eqnarray}

In order to define model C in the presence of structural disorder
we write the static functional
 $\mathcal H$ describing the behaviour of a { disordered} magnetic system
 in equilibrium:
\begin{eqnarray}\label{hamilt1}
{\mathcal H}&=&\int d^d x \Big \{
 \frac{1}{2}\mathring{\tilde{r}} |\vec{\varphi}_0|^2 + V(x)|\vec{\varphi}_0|^2
+\frac{1}{2}\sum_{i=1}^{n}(\nabla \varphi_{i,0})^2 +\nonumber\\
&&\frac{\mathring{\tilde{u}}}{4!}|\vec{\varphi}_0|^4+
\frac{1}{2}a_{{m}}{{m_0}}^2+ \frac{1}{2}\mathring{\gamma}_{{m}}
{{m_0}}|\vec{\varphi}_0|^2-\mathring{h}_{{m}}{{m_0}}
 \Big \},
\end{eqnarray}
where $V(x)$ is an impurity potential which introduces disorder to
the system, and $d$ is the spatial dimension. The functional
(\ref{hamilt1}) contains a coupling $\mathring{\gamma}_m$ to the
secondary density which can be integrated out. Thus  static
critical properties described by the functional (\ref{hamilt1})
are equivalent  to those of a Ginzburg-Landau-Wilson (GLW)
functional:
\begin{eqnarray}\label{hamilt2}
{\mathcal H}=\int d^d x \Big \{
 \frac{1}{2}\mathring{r} |\vec{\varphi}_0|^2{+} V(x)|\vec{\varphi}_0|^2
{+}\frac{1}{2}\sum_{i{=}1}^{n}(\nabla \varphi_{i,0})^2{+}
\frac{\mathring{u}}{4!}|\vec{\varphi}_0|^4
 \Big \}.
\end{eqnarray}
 The parameters $\mathring{r}$ and $\mathring{u}$ are related to
$ \mathring{\tilde r}$, $\mathring{\tilde u}$, $a_m$,
$\mathring{\gamma}_m$ and $\mathring{h}_m$ by
 \begin{eqnarray}\label{relat}
\mathring{r}&=&\mathring{\tilde{r}}+\frac{\mathring{\gamma}_{{m}}\mathring{h}_{{m}}}{a_m},
\qquad\mathring{u}=\mathring{\tilde{u}}-3\frac{\mathring{\gamma}_{{m}}^2}{a_m},
\end{eqnarray}
$\mathring{r}$ is proportional to the temperature distance from
the mean field critical temperature, $\mathring{u}$ is positive.

The properties of the random potential $V(x)$ are governed by a
Gaussian distribution
\begin{equation}\label{r_dist}
{\mathcal P}_{V}={\mathcal N}_{V}\exp\left(-\frac{\int d^d x
{V(x)^2}}{8\mathring{\Delta}}\right).
\end{equation}
with positive width $\mathring{\Delta}$, which is proportional to
the concentration of non-magnetic impurities.

For investigation of statics in the further procedure one has to
take into account that the disorder is quenched. Averaging the
free energy of a disordered system over the distribution
(\ref{r_dist}) with application of the replica trick
\cite{Emery75} one ends up with an effective static functional
containing new terms determined by disorder \cite{Grinstein78}.
Then critical properties are studied on the basis of long-distance
properties of the effective functional.

The procedure for the critical dynamics may be different. We will
treat the critical dynamics of the disordered models within the
field theoretical RG method based on the Bausch-Janssen-Wagner
approach \cite{Bausch76}, where the appropriate Lagrangians of the
models are studied. Therefore we have to obtain the Lagrangian for
our model on the basis of the model equations
(\ref{eq_mov2})-(\ref{hamilt1}). Then after averaging over the
random potential (\ref{r_dist}) we get new terms determined by the
disorder in the Lagrangian. In this case it is not necessary to
apply the replica method \cite{DeDominicis78}. Results can be
written in the form (for details see  \ref{A}) ${\mathcal
L}={\mathcal L}_0+{\mathcal L}_{int}$ with the Gaussian
(unperturbed) part:
\begin{eqnarray}\label{L0}
\fl {\mathcal L}_0=\int d^d x dt \Bigg\{
{-}\mathring{\Gamma}\sum_{i=1}^n{\tilde{\varphi}_{0,i}}{\tilde{\varphi}_{0,i}}{+}
\sum_{i=1}^n{\tilde{\varphi}_{0,i}}
\left({\frac{\partial}{\partial t}}
{+}\mathring{\Gamma}(\mathring{r}{-} \nabla^2)\right)\varphi_{0,i}
\nonumber\\+ \mathring{\lambda}_m
\tilde{m}_0\nabla^2\tilde{m}_0+\tilde{m}_0\left({\frac{\partial}{\partial
t}}-a_m\mathring{\lambda}_m\nabla^2\right)m_0 \Bigg\},
\end{eqnarray}
 and an interaction part:
\begin{eqnarray}\label{Lint}
\fl{\mathcal L}_{int}{=}\int d^d x dt \sum_{i}\Bigg\{
\frac{1}{3!}\mathring{\Gamma}\mathring{u}\tilde{\varphi}_{0,i}
{\varphi}_{0,i}\sum_j{\varphi}_{0,j} {\varphi}_{0,j}{-} \int\!
dt'\! \sum_{j}\!2\mathring{\Gamma}^2{\mathring{\Delta}}
\tilde{\varphi}_{0,i}(x,t) {\varphi}_{0,i}(x,t)\nonumber\\
\times\tilde{\varphi}_{0,j}(x,t') {\varphi}_{0,j}(x,t')+
\mathring{\Gamma}\mathring{\gamma}_m m_0\tilde{\varphi}_{0,i}
{\varphi}_{0,i}{-}
\frac{1}{2}\mathring{\lambda}_m\mathring{\gamma}_m
\tilde{m}_0\nabla^2{\varphi}_{0,i} {\varphi}_{0,i}\Bigg\} \, ,
\end{eqnarray}
with new auxiliary response fields $\tilde{\varphi}_{0,i}$. The
ratio $\mathring u/\mathring\Delta$ determines the degree of
disorder in the system.

Investigating the long-distance and long-time properties of the
theory with the effective Lagrangian ${\mathcal L}$ we  apply
Feynman diagram techniques in order to get dynamical vertex
functions. Details of the calculation of the dynamic vertex
function $\mathring{\Gamma}_{\tilde\varphi,\varphi}$ are given in
\ref{A}. The renormalization of vertex functions leads to the RG
functions, describing the critical dynamics of our model. We use
the minimal subtraction scheme with dimensional regularization to
calculate these functions. General relations are considered in the
next subsections. The details concerning the renormalization
procedure are given in   \ref{B}.

\subsection{RG functions}

From the renormalizing factors introduced in    \ref{B} we define
the $\zeta$-functions which describe the critical behaviour of our
model
\begin{eqnarray}\label{def_z}
\zeta_{a_i}(\{\alpha_i\})&=&-\frac{d\ln Z_{a_i}}{d \ln \mu} \, ,
\end{eqnarray}
here $\{\alpha_i\}=\{u,\Delta,\gamma,\Gamma, \lambda\}$ represents
the set of renormalized parameters for the disordered model C and
$\mu$ is the scale. The notation $a_i$ denotes any density
($\varphi$, $\tilde \varphi$, $m$, $\tilde m$) or any model
parameter $\alpha_i$.

A special case is the additive renormalization $A_{\varphi^2}$ of
the specific heat in the GLW-model. It leads to an additional RG
function
\begin{equation}
B_{\varphi^2}(u,\Delta)=\mu^{\varepsilon}Z^2_{\varphi^2}\mu\frac{d}{d
\mu}\left(Z^{-2}_{\varphi^2}\mu^{-\varepsilon}A_{\varphi^2}\right)
\, ,
\end{equation}
which appears in the static and dynamic $\zeta$-functions.

Since in statics the secondary field can be integrated out one
finds relations to the static $Z$-factors appearing in  model A
(see
 \ref{B}). The relations  (\ref{zfaktorm}) and
(\ref{zfaktorgamma}) lead then to
\begin{equation} \label{first}
\zeta_{\gamma}(u,\Delta,\gamma)=2\zeta_m(u,\Delta,\gamma)+
\zeta_{\varphi}(u,\Delta)+\zeta_{\varphi^2}(u,\Delta),
\end{equation}
and
\begin{equation}\label{zet_m}
\zeta_{m}(u,\Delta,\gamma)=\frac{1}{2}\gamma^2B_{\varphi^2}(u,\Delta).
\end{equation}
Eliminating $\zeta_m$ by inserting Eq. (\ref{zet_m}) into Eq.
(\ref{first}) one gets:
\begin{equation}\label{zet_gg}
\zeta_{\gamma}(u,\Delta,\gamma)=\gamma^2B_{\varphi^2}(u,\Delta)+
\zeta_{\varphi}(u,\Delta)+\zeta_{\varphi^2}(u,\Delta).
\end{equation}

The $\zeta$-functions for the kinetic coefficients $\Gamma$ and
$\lambda$, which are obtained by substituting relations
(\ref{relatZg}) and (\ref{relatZl}) into the definition
(\ref{def_z}), read
\begin{eqnarray}\label{zet_g}
\zeta_{\Gamma}(\{u,\Delta,\gamma,w\}){=}{-}\frac{1}{2}\zeta_{\tilde\varphi}
(\{u,\Delta,\gamma,w\}){+} \frac{1}{2}\zeta_{\varphi}(u,\Delta),
\\ \label{zet_l}
\zeta_{\lambda}(u,\gamma){=}\gamma^2B_{\varphi^2}(u,\Delta).
\end{eqnarray}
The last relation is obtained taking into account Eq.
(\ref{zet_m}).

In order to investigate the dynamical fixed points of model C it
is convenient to introduce the time scale ratio
$
w={\Gamma}/{\lambda} $. The corresponding $\zeta$-function for
this dynamical parameter using the relations (\ref{zet_g}) and
(\ref{zet_l}) reads:
 \begin{eqnarray}\label{zet_w}
\zeta_{w}(u,\Delta,\gamma,w)= \frac{1}{2}\zeta_{\varphi}(u,\Delta)
-\frac{1}{2}\zeta_{\tilde\varphi}(u,\Delta,\gamma,w) -
\gamma^2B_{\varphi^2}(u,\Delta) \, .
\end{eqnarray}

The static critical properties of our model are described by the
flow equations:
\begin{eqnarray}\label{ddu}
l\frac{du}{dl}=\beta_u(u,\Delta), \quad
 \label{ddd}
l\frac{d\Delta}{dl}=\beta_{\Delta}(u,\Delta), \quad
 \label{ddg}
l\frac{d\gamma}{dl}=\beta_{\gamma}(u,\Delta,\gamma),
\end{eqnarray}
with the flow parameter $\ell$ and $\beta$-functions  generally
defined as
 \begin{equation}\label{bet}
\beta_{\alpha_i}(\{\alpha_i\})=\alpha_i[-c_i+\zeta_{\alpha_i}(\{\alpha_i\})],
\end{equation}
where $c_i$ are the engineering dimensions of the corresponding
parameters $\alpha_i$. Following Eq. (\ref{bet}) we can rewrite
the static $\beta$-functions:
\begin{eqnarray}\label{u}
\beta_{u}({u,\Delta})=u\left(-{\varepsilon}-
2\zeta_{\varphi}(u,\Delta)+\zeta_{u}(u,\Delta)\right),
\\ \label{d}
\beta_{\Delta}({u,\Delta})=\Delta\left(-{\varepsilon}-
2\zeta_{\varphi}(u,\Delta)+\zeta_{\Delta}(u,\Delta)\right),
\\
\label{betag}
\beta_{\gamma}({u,\Delta,\gamma})=\gamma\Bigg(-\frac{\varepsilon}{2}-
\zeta_{\varphi}(u,\Delta)-\zeta_{m}(u,\Delta,\gamma)+
\zeta_{\gamma}(u,\Delta,\gamma)\Bigg).
\end{eqnarray}
Note that the flow equations for $u$ and $\Delta$ are independent
from those for the other model parameters. They are the same as
for the diluted $n$-vector model.

The dynamical critical properties are determined by the flow
equation for the time scale ratio $w$, which reads
\begin{eqnarray}
\label{floww} l\frac{dw}{dl}&=&\beta_{w}(u,\Delta,\gamma,w),
\end{eqnarray}
where the $\beta$-function for  $w$, according to Eq. (\ref{bet}),
is defined as
\begin{equation}\label{betaw}
\beta_{w}({u,\Delta,\gamma,w})=w\zeta_{w}(u,\Delta,\gamma,w),
\end{equation}
since $\Gamma$, $\lambda$  and their ratio $w$ are dimensionless
parameters.

Using the relations between $\zeta$-functions for $\beta_{\gamma}$
and $\beta_w$ one obtains:
\begin{eqnarray}\label{g}
\beta_{\gamma}({u,\Delta,\gamma})&=&\gamma\left(-\frac{\varepsilon}{2}+
\zeta_{\varphi^2}(u,\Delta)+\frac{1}{2}{\gamma^2}B_{\varphi^2}(u,\Delta)\right)\nonumber\\
&=&\gamma\left(-\frac{{\cal
A}(u,\Delta)}{2}+\frac{1}{2}{\gamma^2}B_{\varphi^2}(u,\Delta)\right).
\end{eqnarray}
\begin{eqnarray}\label{w}
\beta_{w}({u,\Delta,\gamma,w})=w\Bigg(\frac{1}{2}\zeta_{\varphi}(u,\Delta)
-\frac{1}{2}\zeta_{\tilde\varphi}(u,\Delta,\gamma,w) &-&
\gamma^2B_{\varphi^2}(u,\Delta)\Bigg).
\end{eqnarray}
where  ${\cal A}=\varepsilon-2\zeta_{\varphi^2}(u,\Delta)$ equals
 the ratio of coupling dependent functions of the critical
exponents of the heat capacity $\alpha(u,\Delta)$ and the
correlation length $\nu(u,\Delta)$.

\subsection{Asymptotic properties}

The common zeroes of the $\beta$-functions
(\ref{u}),(\ref{d}),(\ref{g}),(\ref{w}) define the fixed point
(FP) values: $\{\alpha^*\}=\{u^*,\Delta^*,\gamma^*,w^*\}$. The
zeroes of the functions $\beta_u$ and $\beta_{\Delta}$ can be
obtained independent from the other $\beta$-functions. For each
pair of FPs $\{u^*,\Delta^*\}$  one  obtains two values of
$\gamma^*$ from $\beta_{\gamma}$:
 \begin{equation} {\gamma^*}^2{=}0
\quad {\rm and} \quad {\gamma^*}^2{=}\frac{{\cal
A}(u^*,\Delta^*)}{B_{\varphi^2}(u^*,\Delta^*)}{=}\frac{\alpha}{\nu
B_{\varphi^2}(u^*,\Delta^*)},
\end{equation}
where $\alpha$ and $\nu$ are static heat capacity and correlation
length critical exponent calculated at the corresponding FP
$\{u^*,\Delta^*\}$. Then the results for the static FPs are
inserted into $\beta_{w}$ in order to find the corresponding
values of $w^*$.

The relevant FP corresponding to a critical point of the system
has to be (i) accessible from the physical initial conditions and
(ii) stable. The stability of a FP is defined by the eigenvalues
$\omega_i$ of the matrix ${\partial \beta_j}/{\partial \alpha_i}$.
If the real parts of all $\omega_i$ calculated at some FP
$\{\alpha^*\}$ are positive  then the FP $\{\alpha^*\}$ is stable
and the flow  of the system of differential equations (\ref{ddu})
and (\ref{floww}) is attracted to this FP in the limit $\ell\to
0$.

From the structure of the stability matrix we conclude, that the
stability of any FP with respect to the parameters $\gamma$ and
$w$ is determined solely by the derivatives of the corresponding
$\beta$-functions:
\begin{equation}
\omega_{\gamma}=\frac{\partial\beta_{\gamma}}{\partial
{\gamma}},\qquad \omega_{w}=\frac{\partial\beta_{w}}{\partial
{w}}.
\end{equation}
Moreover using (\ref{g}) we can write:
\begin{equation}
\omega_{\gamma}=-\frac{{\cal
A}(u,\Delta)}{2}+\frac{3}{2}\gamma^{2}B_{\varphi^2}(u,\Delta) \, ,
\end{equation}
which at the FP ${\alpha^*_i}$ leads to:
\begin{eqnarray}
\left.\omega_{\gamma}\right|_{{\alpha_i}={\alpha^*_i}}&=&-\frac{\alpha}{2\nu}
\qquad {\rm for} \qquad  {\gamma^*}^2=0 \, ,\\
\left.\omega_{\gamma}\right|_{{\alpha_i}={\alpha^*_i}}&=&\frac{\alpha}{\nu}
\qquad {\rm for} \qquad {\gamma^*}^2\neq 0 \, .
 \end{eqnarray}
 Therefore stability with respect to the parameter $\gamma$
is determined by the sign of $\alpha$.
 For a system with non-diverging heat capacity ($\alpha<0$) at
 the critical point, $\gamma^*=0$ is the stable FP.
 Diluted magnets we consider here always have $\alpha<0$ \cite{review}.
This leads to the  conclusion that in the asymptotic region the
secondary density
 decouples from the OP \cite{Krey76,Krey77,Lawrie84}.

The critical exponents are defined by the FP values of the
$\zeta$-functions. For instance, the asymptotic dynamical critical
exponent $z$  (at the stable FP) is expressed in the following
way:
\begin{equation}\label{zzz}
z=2+\zeta_{\Gamma}(u^*,\Delta^*,\gamma^*,w^*),
\end{equation}
while its effective counterpart in the non-asymptotic region is
defined by the solution of flow equations (\ref{ddu}) and
(\ref{floww}) as
\begin{equation}\label{zzz_eff}
z_{\rm
eff}=2+\zeta_{\Gamma}(u(\ell),\Delta(\ell),\gamma(\ell),w(\ell)).
\end{equation}
In the limit $\ell\to 0$ the effective  exponents reach their
asymptotic values.

\section{Results in one-loop order} \label{III}

Although the one-loop order results have the drawbacks mentioned
in the introduction one gets a qualitatively correct picture of
the effects of disorder on the dynamics. Therefore we discuss this
case in more detail.

We are interested in the dynamical properties of disordered model
C within the first nontrivial order of expansion in renormalized
couplings, that is the one-loop order. The static functions
$\beta_u$ and $\beta_{\Delta}$ for disordered systems are known in
higher loop approximation \cite{Folk00}. However taking the same
order for statics as for dynamics, we restrict the expressions for
$\beta_u$ and $\beta_{\Delta}$  to the one-loop approximation:
\begin{eqnarray}\label{b1}
  \beta_{1u}&=&u\left(-{\varepsilon}+\frac{n+8}{6}u-{24\Delta}\right),\\
  \label{b2}
  \beta_{1\Delta}&=&\Delta\left(-{\varepsilon}-{16\Delta}+\frac{n+2}{3}u\right).
\end{eqnarray}
Note that the region of physical relevance of the couplings
$u,\Delta$ for diluted magnets is restricted by $u\geq 0$,
$\Delta\geq 0$. The other $\beta_{\gamma}$ is obtained using the
static one-loop $\zeta$-function:
\begin{eqnarray}\label{b3}
 \beta_{1\gamma}&=&
\gamma\left(-\frac{\varepsilon}{2}-{4\Delta}+\frac{n+2}{6}u+\frac{1}{2}\frac{\gamma^2n}{2}\right),
\end{eqnarray}
 For obtaining the dynamic function
 $\beta_{w}$, we should first calculate the renormalizing factor
 $Z_{\tilde\varphi}$. From the one-loop part of  $Z_{\tilde\varphi}$ (see \ref{A})
following formula (\ref{def_z}) we can obtain the
$\zeta_{\tilde{\varphi}}$-function. Using obtained result together
with one-loop static $\zeta$-functions we get $\beta_{w}$ in the
following form:
\begin{eqnarray}
\beta_{w}&=& w\left({4\Delta}+\gamma^2\frac{w}{1+w}-
\frac{\gamma^2n}{2}\right).\label{b4}
\end{eqnarray}

\begin{table}[tbp]
\caption{\label{fpstab1}One-loop FPs  of model  C with disorder
and their stability for $\varepsilon=4-d>0$. Note that
${\gamma^*}^2$ has to be positive for the existence of a real FP.
.}
\begin{indented} \item[]
 \begin{tabular}{|c|c|c|c|c|c|}
 \hline
FP&$u^*$ & $\Delta^*$& ${\gamma^*}^2$& $\rho^*$ &stability\\
 \hline
 {\bf G}&0&0&0&$0\leq \rho^* \leq 1$&unstable\\
 \cline{1-1} \cline{4-6}&&&      &0 &\\
{\bf G$^\prime$} &&&$\frac{2}{n}\varepsilon$& $n/2$ &unstable\\
 \cline{5-5}&&&      &1 &\\
 \hline
 {\bf P}&$\frac{6\varepsilon}{n+8}$&0&0&$0\leq \rho^* \leq 1$&marginal for $n>4$\\
\cline{1-1} \cline{4-6}&&&      &0 &\\\cline{5-5}
 {\bf
P$^\prime$}&&&$\frac{2}{n}\frac{(4-n)\varepsilon}{n+8}$&$n/2$
&unstable\\
 \cline{5-5}&&&      &1 &\\
 \hline {\bf U}&0&$-\frac{3\varepsilon}{4}$&0&0&unstable\\
 \cline{5-5}&&&      &1 &\\
 \cline{1-1}\cline{4-6}&&&      &0 &unstable\\  \cline{5-6} {\bf
U$^\prime$}&&&$\frac{\varepsilon}{n}$& $3n/4$ &stable for
$n<4/3$\\
 \cline{5-6}&&&      &1 &stable for $n>4/3$\\
 \hline {\bf
M}&$\frac{3\varepsilon}{2(n-1)}$&$\frac{(4-n)\varepsilon}{32(n-1)}$&0&0&stable
for $1<n<4$\\ \cline{5-6}&&&      &1 &unstable\\
 \cline{1-1}\cline{4-6}&&&      &0 &unstable\\  \cline{5-5} {\bf
M$^\prime$}&&&$\frac{2}{n}\frac{n-4}{4(n-1)}\varepsilon$&$3n/4$&ustable\\
\cline{5-6}&&&      &1 &unstable\\
 \hline
 \end{tabular}
 \end{indented}
   \end{table}

Setting the right hand side of Eqs. (\ref{b1})-(\ref{b4}) equal to
zero we obtain the system of equations for the FPs.  The structure
of the functions $\beta_u$ and $\beta_{\Delta}$ leads to the
existence of four FPs: the Gaussian FP {\bf G} $\{ u^*=0,
\Delta^*=0 \}$ , the 'polymer' FP {\bf U} $\{u^*=0, \Delta^* \neq
0\}$, the FP of the pure system {\bf P} $\{ u^*\neq 0, \Delta^*=0
\}$ and the mixed FP {\bf M} $\{u^*\neq 0, \Delta^*\neq 0\}$.
Depending on the FP values $u^*$ and $\Delta^*$  values for the
genuine model C parameters can be found. As it was already
mentioned two values ${\gamma^*}^2=0$ or ${\gamma^*}^2\neq 0$
correspond to each FP, this doubles the number of fixed points,
see Table \ref{fpstab1}. We note that the FPs with nonzero
${\gamma^*}^2$ are indicated by a prime $^\prime$. We do not
consider further the FPs {\bf U} and {\bf U$^\prime$} since these
FPs lie outside the physical region of positive values of
$\Delta$.

 Among the rest of the
FPs only one is stable depending on the OP dimension $n$. For
$n>4$ FP {\bf P} is stable while  for $1<n<4$ FP {\bf M} is
stable. Thus the one-loop value of marginal dimension
$n_c(\varepsilon)=4$  defines the borderline where $\alpha=0$. In
higher-loop order the FPs picture is not changed apart from a
change of the borderline function $n_c(\varepsilon)$. Estimates
obtained on the basis of six-loop order calculations
\cite{Bervillier86} give
 definitely $n_c<2$ at $d=3$.

In order to discuss the dynamical FPs it turns out to be useful to
introduce the parameter $\rho=w/(1+w)$ which maps  $w$ and its FPs
 on a finite region of the parameter
space. Then instead of the flow equation (\ref{floww}) the flow
equation for $\rho$ arises:
\begin{equation}\label{drho}
l\frac{d\rho}{dl}=\beta_{\rho}(u,\gamma,\rho),
\end{equation}
where according to (\ref{betaw})
\begin{equation}\label{brho}
\beta_{\rho}(u,\gamma,\rho)=\rho(\rho-1)\zeta_w(u,\gamma,\rho) \,
.
\end{equation}

Setting the right side of (\ref{brho}) to zero and using the FP
values from Table~\ref{fpstab1} the values of $\rho^*$ can be
found. They are shown in Table~\ref{fpstab1} as well.

 Each non-zero solution of
${\gamma^*}^2$  leads according to (\ref{drho}) to three dynamical
FPs: either with $\rho^*=0$ (i.e. $w^*{=}0$), $\rho^*{=}1$ (i.e.
$w^*{=}\infty$) or $0<\rho^*<1$ correspondingly. For the FP which
corresponds to ${\gamma^*}^2=0$ the situation is the following.
For the FP {\bf M} only two dynamical FPs with $\rho^*=0$ and
$\rho^*=1$ exist, while for the  FPs {\bf P} and {\bf G} any value
of $\rho^*$ between 0 and 1 is allowed. Checking the stability of
these FPs we see that for $1<n<4$ only FP {\bf M} with $\rho^*=0$
is stable. The corresponding one-loop asymptotic value of the
dynamical critical exponent  in this case coincides with the
one-loop result for the pure relaxational model (model A) with
disorder: $z=2+\frac{(4-n)\varepsilon}{8(n-1)}$
\cite{Grinstein77,Krey77}. For $n>4$  FP {\bf P} is marginal, all
other FPs are unstable. Formally flow starting from the initial
values with $\rho_i$ ends up at some FP {\bf P} with
$\rho^*(\rho_i)$ depending on the initial value $\rho_i$. Thus one
gets a whole line of FPs at $u^*\not=0$, $\Delta^*=\gamma^*=0$. In
this case the one-loop result for $z$ coincides with the
conventional theory value $z=2$. Since in one-loop approximation
$n_c(\varepsilon)=4$ the FPs {\bf M} and {\bf P} determine the
critical behaviour of the disordered model C for $n<4$ and $n>4$
respectively.

\section{Two-loop results}\label{IV}

Static RG functions are already known in high-order approximations
within different renormalization schemes (for references see e. g.
Ref \cite{review}). Two-loop expressions for functions $\beta_u$,
$\beta_{\Delta}$, $\beta_{\gamma}$ within the minimal subtraction
scheme can be obtained in the replica limit from results of Ref.
\cite{Kyriakidis96} and they read:
\begin{eqnarray}
\fl \beta_{u}=\beta_{1u}+u\Big( -\frac{3m+14}{12}u^2
+\frac{22m+116}{3}u\Delta-328\Delta^2\Big),\\ \fl
\beta_{\Delta}=\beta_{1\Delta}+\Delta\Big( -5\frac{m+2}{36}u^2
 +22\frac{m+2}{3}u\Delta-168\Delta^2\Big),\\
\fl \beta_{\gamma}=\beta_{1\gamma}+\gamma\Big(
-\frac{5(m+2)}{72}u^2
 +\frac{5(m+2)}{3}u\Delta
 -20\Delta^2\Big).\label{gg}
 \end{eqnarray}

The dynamical function $\beta_{\rho}$ is given by formulas
(\ref{zet_w}) and (\ref{brho}), where functions $\zeta_{\tilde
\varphi}$ and $\zeta_{\varphi}$ are needed also for the
calculation of the dynamical critical exponent $z$,
$B_{\varphi^2}$ is the same as in one-loop approximation.
$\zeta_{\varphi}$ can be obtained in the replica limit from
results of Ref. \cite{Kyriakidis96}:
\begin{equation}
\zeta_{\varphi}=-\frac{m+2}{72}u^2+\frac{m+2}{3}u\Delta-4\Delta^2.
\end{equation}
For the calculation of $\zeta_{\tilde \varphi}$ we use the RG
scheme described in detail in Appendices A and B.  We obtain
$\zeta_{\tilde \varphi}$ from the two-loop value of
$Z_{\tilde\varphi}$ (\ref{zfac}) , according to the definition
(\ref{def_z}) it reads:
 \begin{eqnarray}\label{zpt}
\fl \zeta_{\tilde \varphi}=-8{\Delta}+
 3\frac{n+2}{3}u\Delta-
44\Delta^2-
\frac{n+2}{6}{u}^2\left(\ln\frac{4}{3}-\frac{1}{12}\right)
-2{\gamma^2}\frac{w}{1+w}\nonumber\\+
\Bigg[\frac{n+2}{3}u\left(1-3\ln\frac{4}{3}\right)\nonumber\\+
{\gamma^2}\frac{w}{1{+}w}\Bigg(\frac{n}{2}-
\frac{w}{1{+}w}-\frac{3(n{+}2)}{2}\ln\frac{4}{3} -
\frac{1{+}2w}{1{+}w}\ln\frac{(1{+}w)^2}{1{+}2w}\Bigg)\Bigg]{\gamma^2}\frac{w}{1{+}w}\nonumber\\-
12\Delta{\gamma^2}\frac{w}{1{+}w}+
4{\Delta}{\gamma^2}\frac{w}{1{+}w}\Bigg[{w}\ln\frac{w}{1{+}w}-
 {3}\ln(1+w) -\frac{w}{1{+}w}\ln w\Bigg].
 \end{eqnarray}
Setting the  coupling $\Delta$ equal to zero in (\ref{zpt}), one
recovers the two-loop result for pure model C \cite{note2}, while
expression (\ref{zpt}) with $\gamma=0$ corresponds to the function
$\zeta_{\tilde \varphi}$ of model A with dilution, which was
extensively studied
\cite{Grinstein77,Prudnikov92,Oerding95,Janssen95}. The
$\gamma^2\Delta$-term is an intrinsic contribution of the
disordered model C.

Two complementary ways are known for the analysis of the FP
equations. The first one is the $\varepsilon$-expansion
\cite{Wilson72}, while the second one consists in fixing
$\varepsilon$ and solving equations for the FP numerically
\cite{Schloms87}. As it is known from statics for  diluted
systems, expansions in $\varepsilon$ do not give reliable
numerical estimates for critical exponents. For instance for
disordered Ising magnets the $\varepsilon$-expansion technique
leads in fact to a $\sqrt\varepsilon$-expansion
{\cite{sqrteps,Grinstein78}}, that does not give trustable results
for $\varepsilon=1$ \cite{Folk00}. Therefore we follow the second
way of analysis  working directly in the $d=3$ dimensional space.

The series for RG functions are known to be asymptotic at best.
For instance, no FP of the fourth-order couplings is obtained
without application of resummation for diluted systems in the
two-loop approximation \cite{review}. Therefore for static RG
functions it is standard  to apply resummation technique in order
to obtain reliable results.  We use here the Pad\'e-Borel
resummation procedure \cite{Baker78} for the functions
$\{\beta_u/u, \beta_{\Delta}/\Delta,
\beta_{\gamma}/\gamma-\gamma^2n/4\}$.

\subsection{Fixed points and their stability}
We can analyze the RG functions $\beta_u$ and $\beta_{\Delta}$
independently from the other ones. The results for these static
functions are already known \cite{review}.  The outcome of the two
loop approximation qualitatively repeats the results of the
one-loop approximations, namely,   the Gaussian FP {\bf G}, the FP
of the pure system {\bf P}  and the mixed FP {\bf M} are found.
However, contrary to the one-loop results, the FP {\bf M} is
defined also for the Ising case $n=1$.

Among the remaining FPs only one  is stable depending on the order
parameter dimension $n$. For $n<n_c$, the FP {\bf M} is the stable
one, at $n=n_c$ FPs {\bf M} and {\bf P} change stability, and for
$n>n_c$ FP {\bf P} becomes stable. For the stability boundary at
$d=3$ we find $n_c=1.55$ which is somewhat smaller than
higher-loop order estimations \cite{Bervillier86}.

From the structure of two-loop $\beta_{\gamma}$ (\ref{gg}) one
concludes that all  FPs described above exist for $\gamma^*=0$.
Analyzing $\beta_{\gamma}$ we find that only the Gaussian FP {\bf
G} and the FP {\bf P} have  counterparts with positive non-zero
$\gamma^*$: FP {\bf G$^\prime$} for  all $n$ and FP {\bf
P$^\prime$}  for $n<n_c$ respectively. The FP values of the model
parameters and the dynamical exponents for all FPs are listed in
 Table~\ref{Fps} for the numbers of OP components $n=1,\,2,\,3$.

At FP {\bf G} a line of FP for the time scale ratio $0 \leq \rho
\leq 1$ is obtained  for any $n$. For the  FPs with $\gamma^*=0$
only two values ${\rho}^*=0$ or ${\rho}^*=1$ are found, while for
the FPs {\bf P$^\prime$} and {\bf G$^\prime$} one obtains three
solutions ${\rho}^*=0$, ${\rho}^*=1$ and a non-zero solution with
$0<{\rho}^*<1$. All FPs are given in Table~\ref{Fps}. We denote
the FPs with ${\rho}^*=1$  by subscript $_1$, whereas FPs with
$\rho^*=0$  have no subscript. Since FP {\bf P$^\prime$} with
non-zero solution $0<{\rho}^*<1$ corresponds to the FP of pure
model C,  we denote it by  {\bf C}, and the FP {\bf G$^\prime$}
with non-zero $\rho^*$ we denote  by {\bf G$^\prime_C$}.

Analyzing the stability exponents $\omega_{\gamma}$ and
$\omega_{\rho}$ we see that only FP {\bf M} is stable for $n=1$
and FP {\bf P} is stable for $n=2,3$. That means that the model A
critical behaviour is reached  in the asymptotics in any case
(diluted model A universality class for $n<n_c$ and pure model A
universality class for $n>n_c$). As a consequence the secondary
density is for all $n$ asymptotically decoupled and  formally the
dynamical critical exponent takes its Van Hove value $z_m=2$, as
expected due to arguments of Ref. \cite{Krey76,Krey77}.

 The stability exponents are
listed in  Table~\ref{omeg}. They determine the ``velocity" of the
RG flows in the FPs vicinity. A small value of a stability
exponent calculated at a stable FP means slow approach to this FP
in the corresponding direction. For example, for FP {\bf M} at
$n=1$ one has $\omega_{\gamma}=0.002$, which leads to slow
approach of $\gamma$ to its FP value $\gamma^*=0$.

\begin{table}[htp]
\caption {\label{Fps} Two-loop values for the FPs of disordered
model C.}
\begin{center}
\begin{indented} \item[]
   \begin{tabular}{|c|c|c|c|c|c|c|}
\hline $n$&FP & $u^*$ & $\Delta^*$ & $\gamma^*$&$\rho^*$&$z$\\
   \hline
\hline $\forall n$& {\bf G}&0&0&0& $0\le\rho^*\le 1$&2\\ \hline
 $n=1$ &{\bf G$^\prime$}&0&0&1.4142& 0 &2\\
 &{\bf G$^\prime_C$}&0&0&1.4142&0.3446  &3\\
 &{\bf G$^\prime_1$}&0&0&1.4142& 1 &$\infty$\\
&{\bf P} &1.3146 &0 &0&0&2.052\\ &{\bf P$_1$}  &1.3146 &0 &0
&1&2.052\\
 &{\bf P$^\prime$} &1.3146 &0 &0.4582&0&2.052
\\ &{\bf C} &1.3146 &0 &0.4582 &0.2664&2.105 \\
 &{\bf P$^\prime_1$} &1.3146 &0 &0.4582 &1 &2.052\\
&{\bf M} &1.6330 &0.0209 &0&0&2.139\\   &{\bf M$_1$} &1.6330
&0.0209 &0&1&2.139\\
 \hline
$n=2$ &{\bf G$^\prime$}&0&0&1& 0 &2\\
 &{\bf G$^\prime_C$}&0&0&1&0.6106  &3\\
 &{\bf G$^\prime_1$}&0&0&1& 1 &$\infty$\\
&{\bf P} &1.1415 &0 &0&0&2.053\\  &{\bf P$_1$} &1.1415 &0
&0&1&2.053\\ \hline
 $n=3$&{\bf G$^\prime$}&0&0&0.8165& 0 &2\\
 &{\bf G$^\prime_C$}&0&0&0.8165&0.7993  &3\\
 &{\bf G$^\prime_1$}&0&0&0.8165& 1 &$\infty$\\
 &{\bf P} &1.0016 &0 &0&0&2.051\\   &{\bf P$_1$} &1.0016 &0 &0&1&2.051\\ \hline
 \end{tabular}
 \end{indented}
\end{center}
\end{table}

\begin{table}[htp]
\caption {\label{omeg} Stability exponents for the FPs given in
Table~\protect\ref{Fps}}
\begin{center}
\begin{indented}\item[]
   \begin{tabular}{|c|c|c|c|c|c|}
\hline $n$ &FP&  $\omega_u$ & $\omega_{\Delta}$ &
$\omega_{\gamma}$&$\omega_{\rho}$\\
   \hline
   \hline $\forall n$ &{\bf G}&-1&-1&-0.5&0\\
   \hline
 $n=1$ &{\bf G$^\prime$}&-1&-1&1& -1 \\
 &{\bf G$^\prime_C$}&-1&-1&1&0.976 \\
 &{\bf G$^\prime_1$}&-1&-1&1& -$\infty$ \\
   &{\bf P} &0.566 &-0.105 &-0.053 &0.052\\&{\bf
P$_1$}&0.566 &-0.105 &-0.053 &-0.052\\ &{\bf P$^\prime$}&0.566
&-0.105&0.105&-0.053
\\&{\bf C} &0.566 &-0.105&0.105&0.041 \\
 &{\bf P$^\prime_1$}&0.566 &-0.105&0.105&-$\infty$ \\
  &{\bf M}&0.494 &0.194 &0.002&0.139\\ &{\bf M$_1$} &0.494 &0.194
&0.002&-0.139\\ \hline $n=2$ &{\bf G$^\prime$}&-1&-1&1& -1 \\
 &{\bf G$^\prime_C$}&-1&-1&1&0.745 \\
 &{\bf G$^\prime_1$}&-1&-1&1&-$\infty$ \\ &{\bf P} &0.581 &0.078
&0.039&0.053\\ &{\bf P$_1$}&0.582 &0.078 &0.039&-0.053\\ \hline
  $n=3$
  &{\bf G$^\prime$}&-1&-1&1& -1 \\
 &{\bf G$^\prime_C$}&-1&-1&1&0.522 \\
 &{\bf G$^\prime_1$}&-1&-1&1&-$\infty$\\
  & {\bf P} &0.597 &0.222 &0.111&0.051\\&{\bf P$_1$}&0.597 &0.222 &0.111&-0.051\\ \hline
 \hline
\end{tabular}
\end{indented}
\end{center}
\end{table}

In  Table \ref{Fps} we give the numerical values of the asymptotic
dynamical critical exponents $z$ calculated for all FPs. If the
flow from the initial values of the couplings passes near one of
these FPs one may observe an effective critical behaviour governed
by the values of the critical exponents corresponding to that FP.

\subsection{Flows and effective exponents $z_{\rm eff}$}

The nonasymptotic behaviour is described by the flow of the static
couplings and the dynamic parameter under renormalization. It can
be obtained solving the flow equations (\ref{ddu})
 and (\ref{drho}). Using the solution we
 get
the behaviour ofthe  effective critical exponent $z_{\rm eff}$
with continuous change of the flow parameter.

\begin{figure}[htbp]
\begin{center}
{\includegraphics[width=0.35\textwidth]{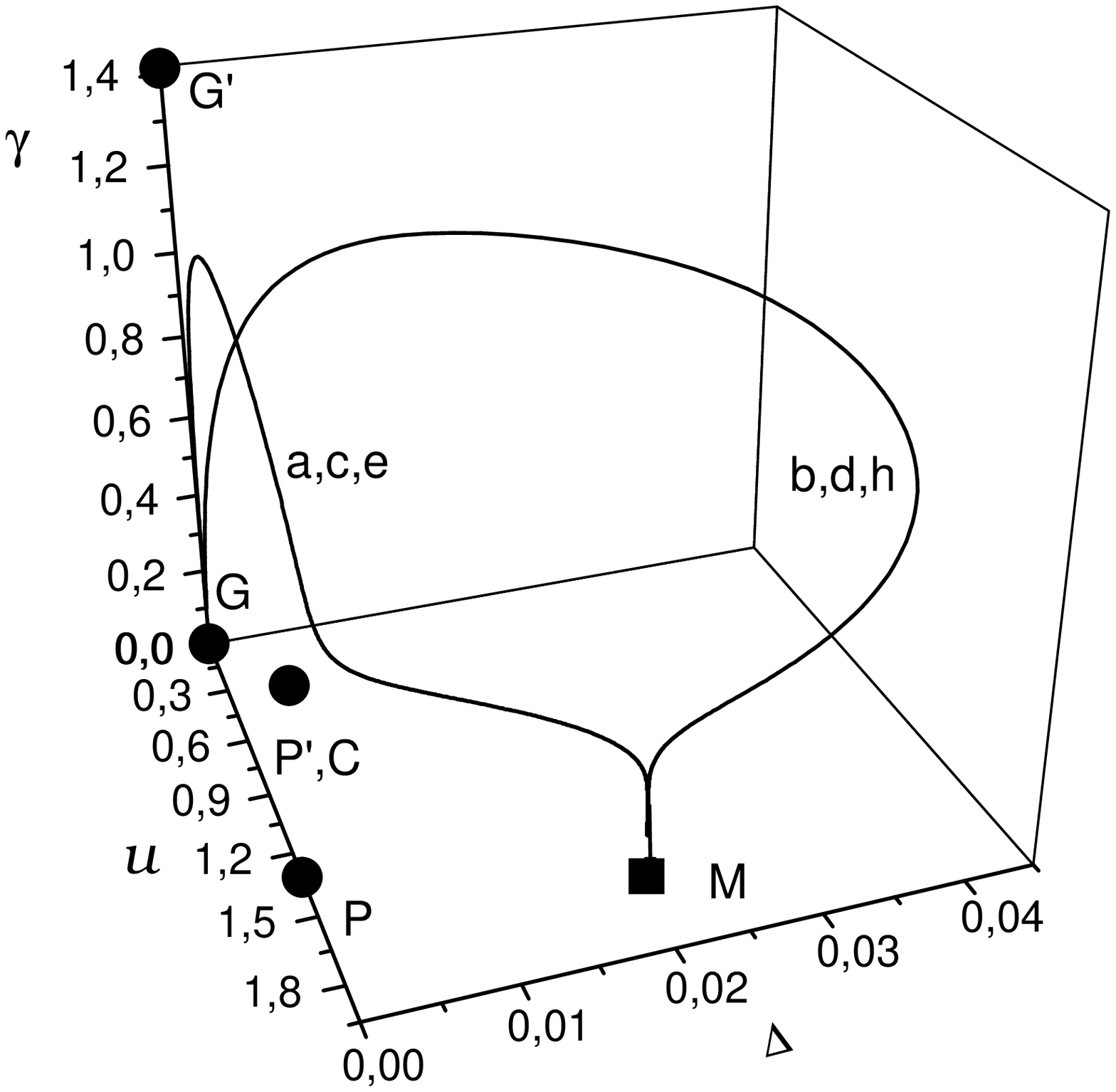}}
{\includegraphics[width=0.35\textwidth]{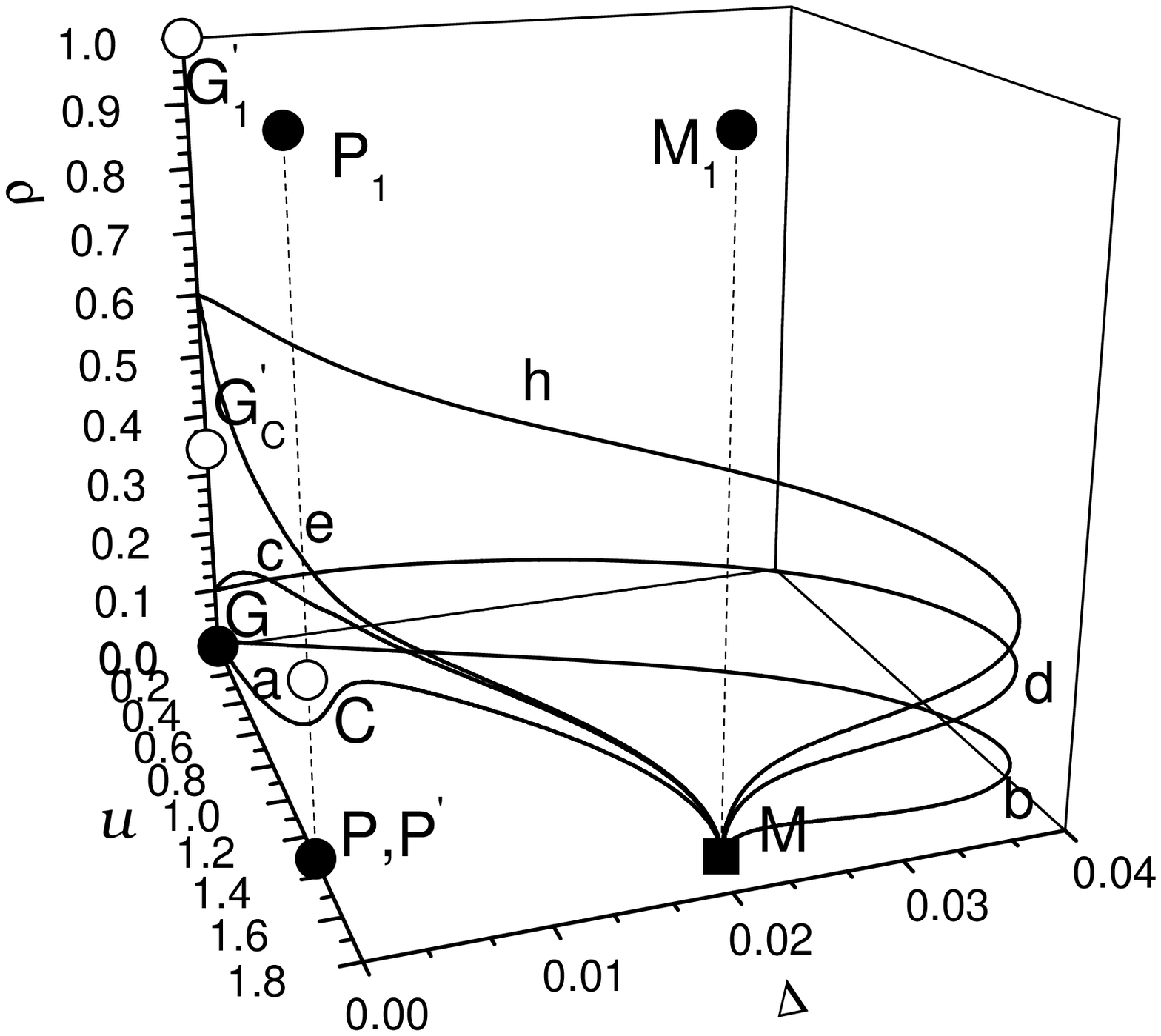}} \end{center}
\caption{\label{dflow} Static (left picture) and dynamic (right
picture) flows of disordered model C for $n=1$ ($n<n_c$). Filled
circles mean unstable FPs with $\gamma^*=0$, blank circles
indicate projections of unstable FPs with nonzero $\gamma^*$,
while the filled square denotes the stable FP.}
\end{figure}

First we consider the case of Ising spins $n=1$. We present static
flows in the subspace $u-\Delta-\gamma$ as well as a ``projection"
of the dynamic flows on the subspace of the model parameters
$u-\Delta-\rho$ in  Fig.~\ref{dflow}. All flows are obtained for
initial values with $\gamma(\ell_0)=0.1$. Flow equation solutions
presented in  Fig.~\ref{dflow} are obtained for different ratios
$\Delta(\ell_0)/u(\ell_0)$ as well as different values of
$\rho(\ell_0)$. Flows starting from initial conditions with small
ratio $\Delta(\ell_0)/u(\ell_0)$ (curves a,c,e) correspond to a
small degree of disorder. They are influenced by the unstable FPs
{\bf P}, {\bf P$^\prime$}, {\bf C}  and {\bf G$^\prime_C$} while
flows corresponding to larger disorder (curves b, d, h)
 are  affected only by the presence of FP {\bf G$^\prime_C$}.

\begin{figure}[htbp]
\begin{center}
{\includegraphics[width=0.35\textwidth]{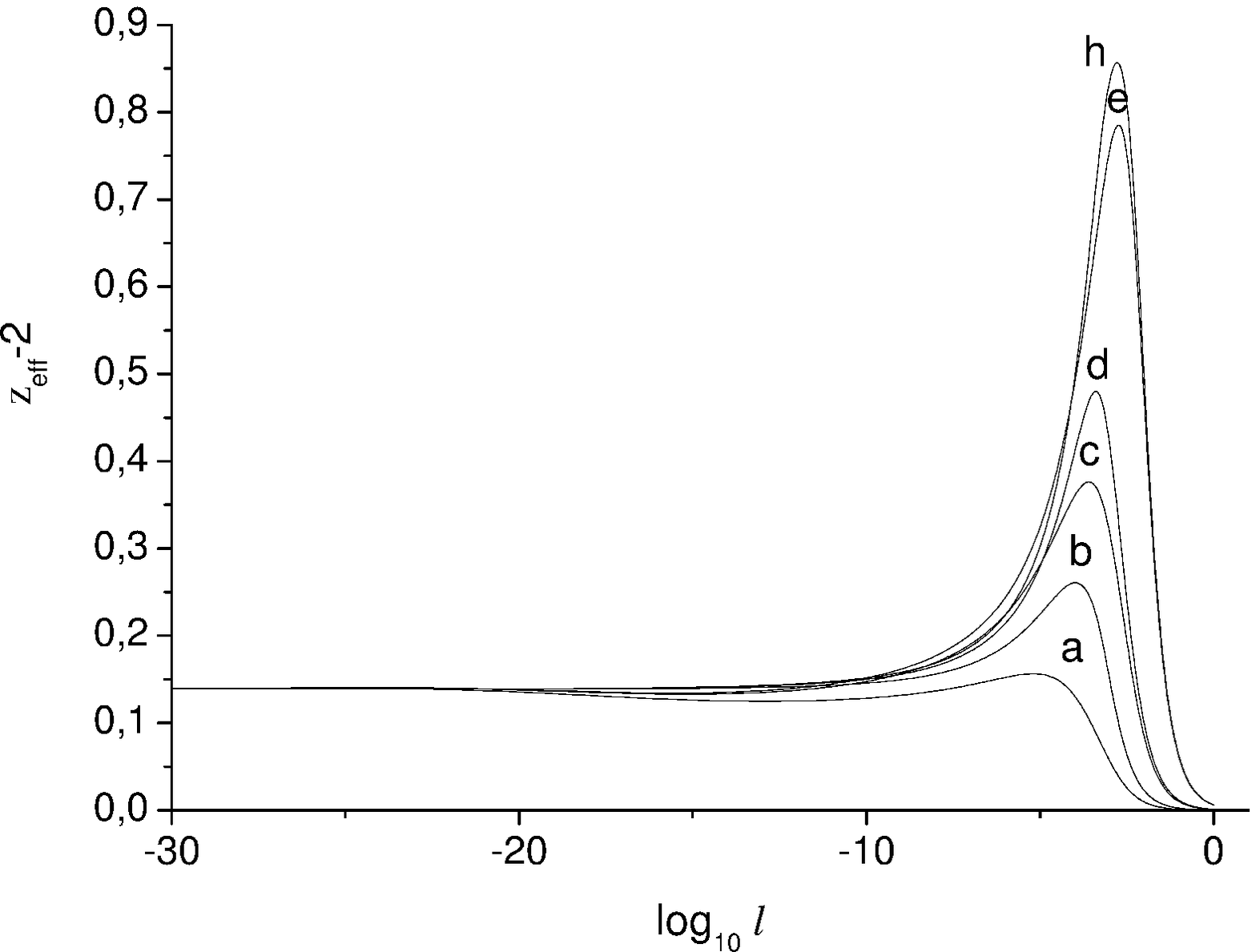}}
{\includegraphics[width=0.35\textwidth]{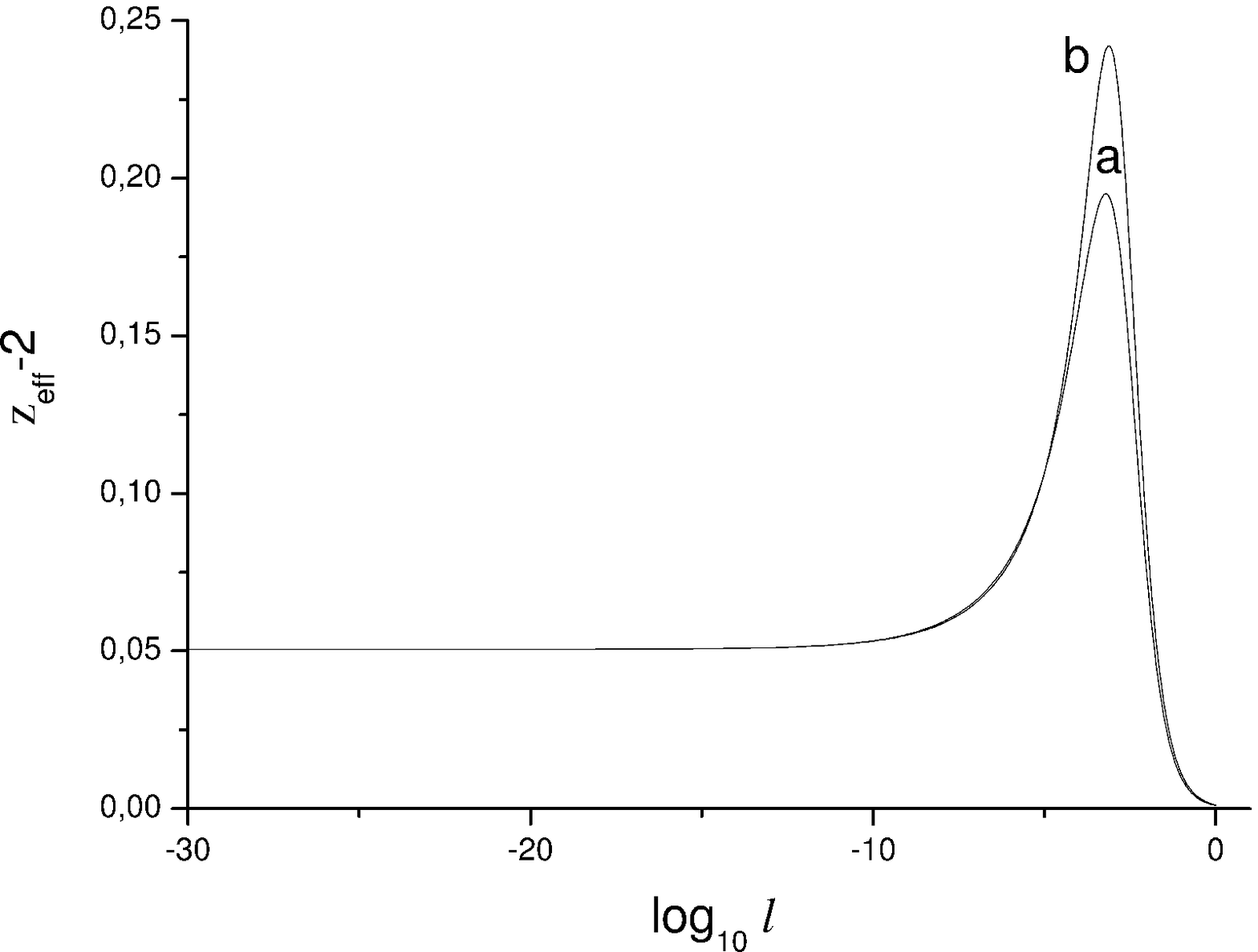}} \end{center}
\caption{\label{z1} Dependencies of the effective dynamical
exponent on the flow parameter for OP dimension $n=1$ and $n=3$.
Curves are obtained on the basis of flows of Fig.~\ref{dflow} and
Fig.~\ref{dflow2}. }
\end{figure}

Fig.~\ref{z1}  presents the flow parameter dependencies of $z_{\rm
eff}$  for the curves in Fig.~\ref{dflow}. The intrinsic feature
of all curves is a non-monotonic behaviour of the effective
dynamical critical exponent with approach to an asymptotic value.
The experimental investigations are performed mainly in the
non-asymptotic region. As it follows from Fig.~\ref{z1} one can
observe  values of the dynamical exponent $z$ that exceed or are
smaller  than an asymptotic one. Our {\em asymptotic} value
$z=2.14$ is somewhat smaller than the central value of the
experimental result $z=2.18\pm0.10$ \cite{Rosov92} obtained for
the dynamics of diluted Ising systems, however this experimental
outcome is in agreement with our {\em non-asymptotic}
observations.

\begin{figure}[htbp]
\begin{center}
{\includegraphics[width=0.35\textwidth]{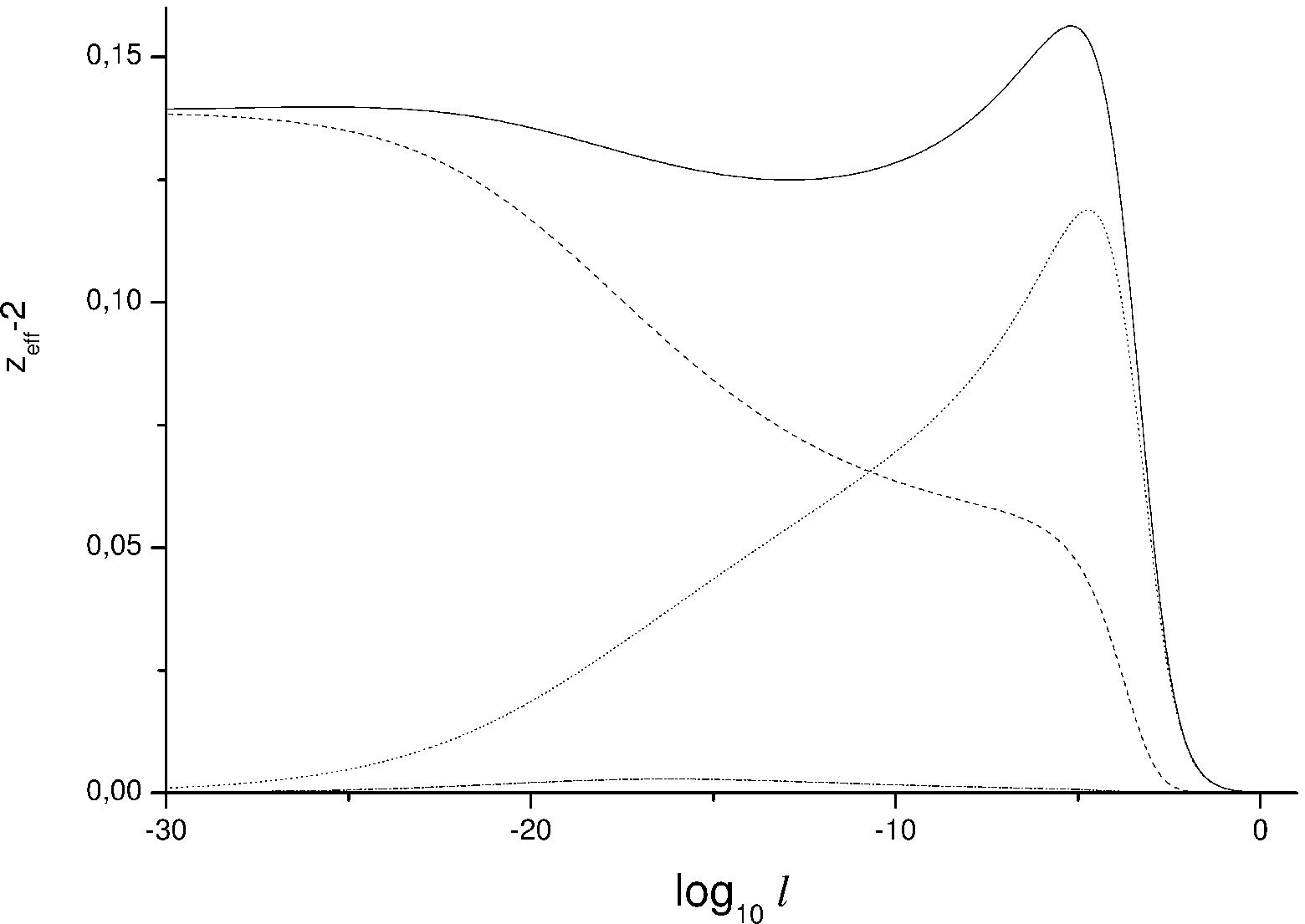}}
{\includegraphics[width=0.35\textwidth]{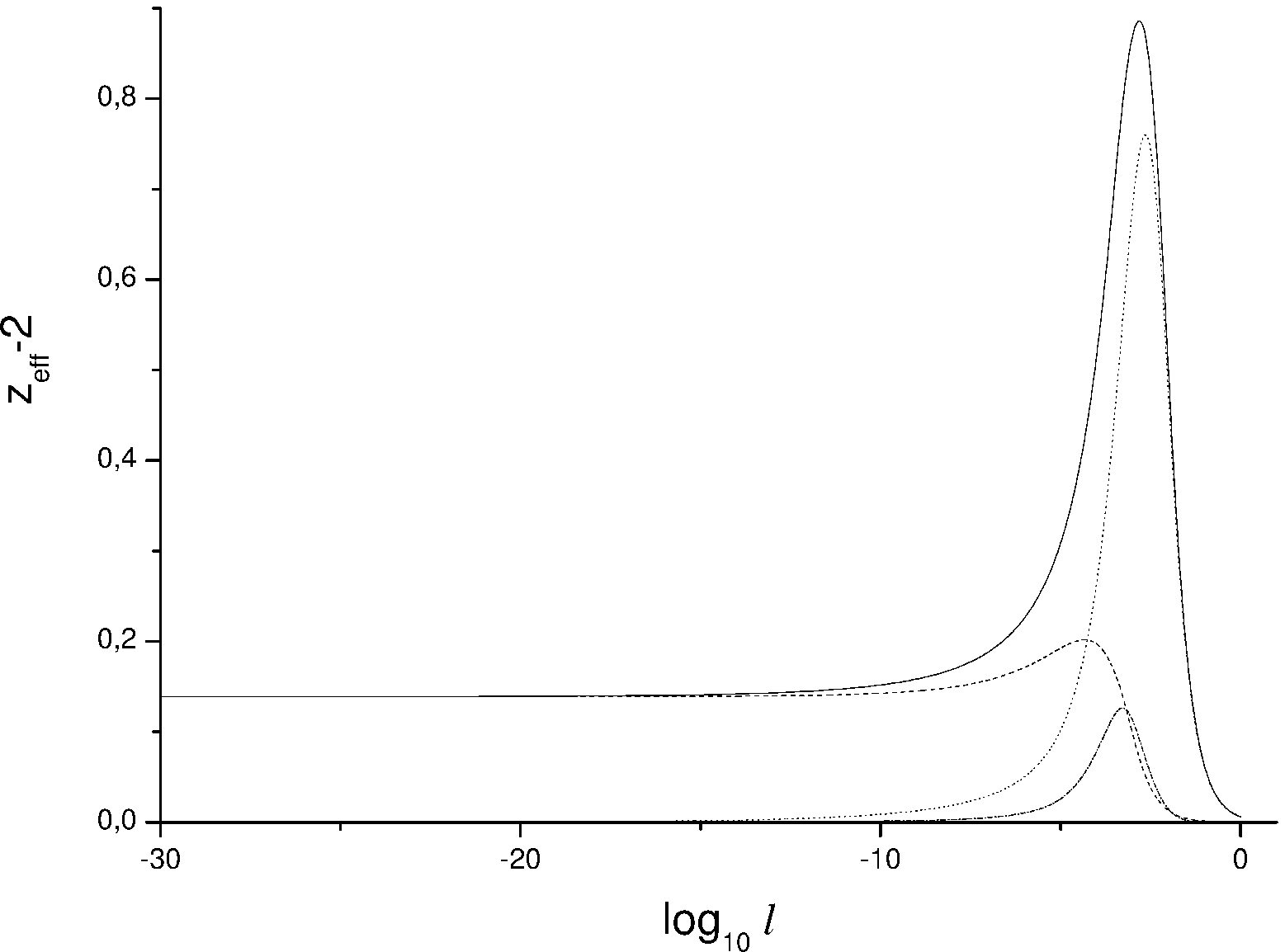}}
\end{center}
\caption{\label{cont1} Different contributions to the effective
dynamical critical exponent corresponding to the flows a and h of
Fig. \protect\ref{dflow}. The solid line represents the complete
$z_{\rm eff}$, for the other lines see the text. }
\end{figure}

It is interesting to look at contributions of different origin to
the effective dynamical critical exponent $z_{\rm eff}$. They are
shown in  Fig. \ref{cont1}   for  flows a and h of Fig.\ref{dflow}
correspondingly. The contributions consist of (i) the terms
already present in model A (dashed curve in Fig. \ref{cont1}),
(ii)  terms present in pure model C only (short dashed curve) and
(iii) terms present in the diluted model C only (dashed-short
dashed curve) \cite{note1}. The interplay of the above
contributions gives the full effective exponent $z_{\rm eff}$
(solid line) and may lead to an almost { asymptotic} value of the
exponent { although} the parameters are far away from asymptotics.
This is an important point, since the appearance of an asymptotic
value in one physical quantity does not mean that other quantities
have also reached the asymptotics. This is due to the different
dependence of physical quantities on the model parameters.

\begin{figure}[htbp]
\begin{center}
{\includegraphics[width=0.35\textwidth]{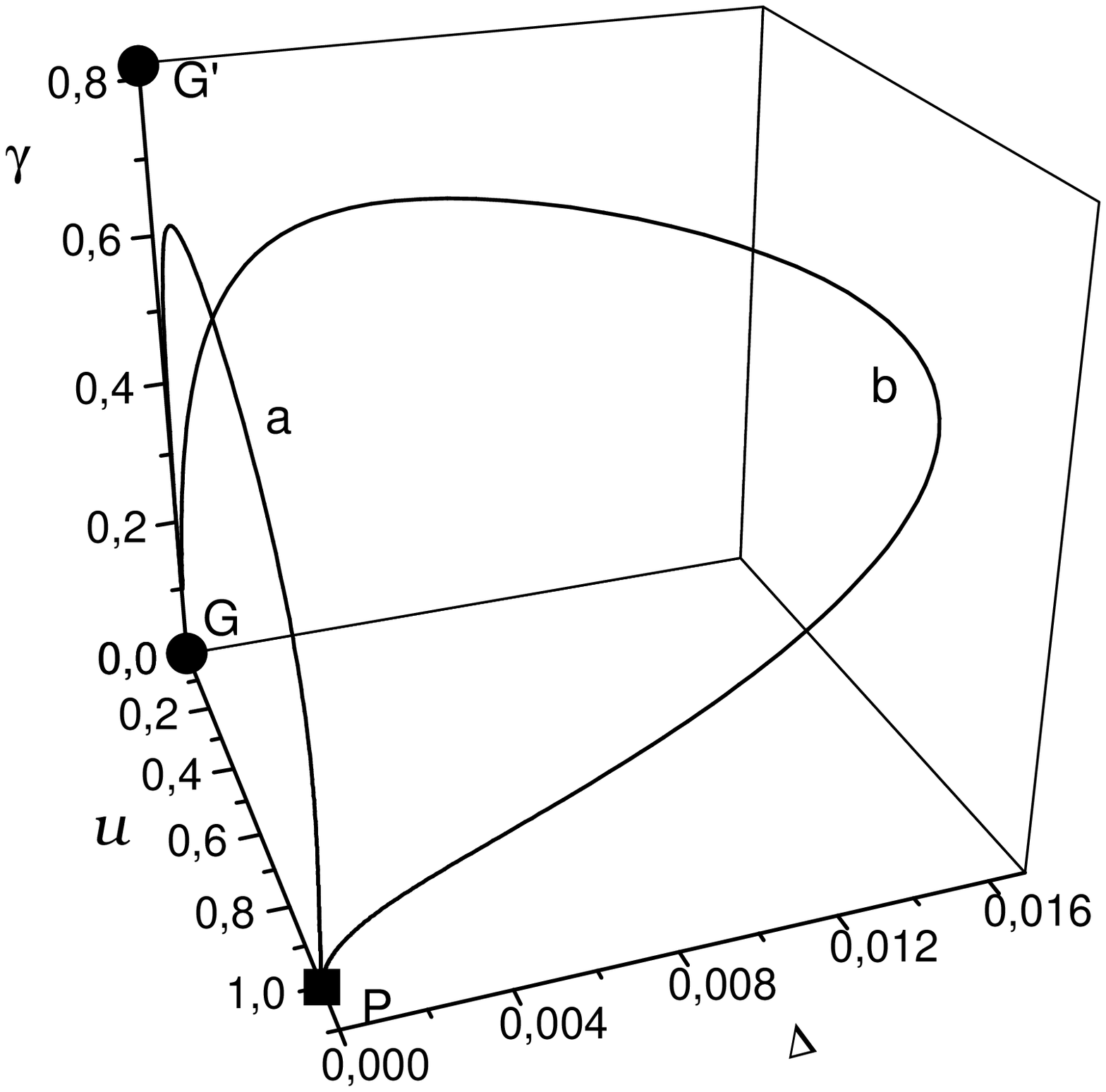}}
{\includegraphics[width=0.35\textwidth]{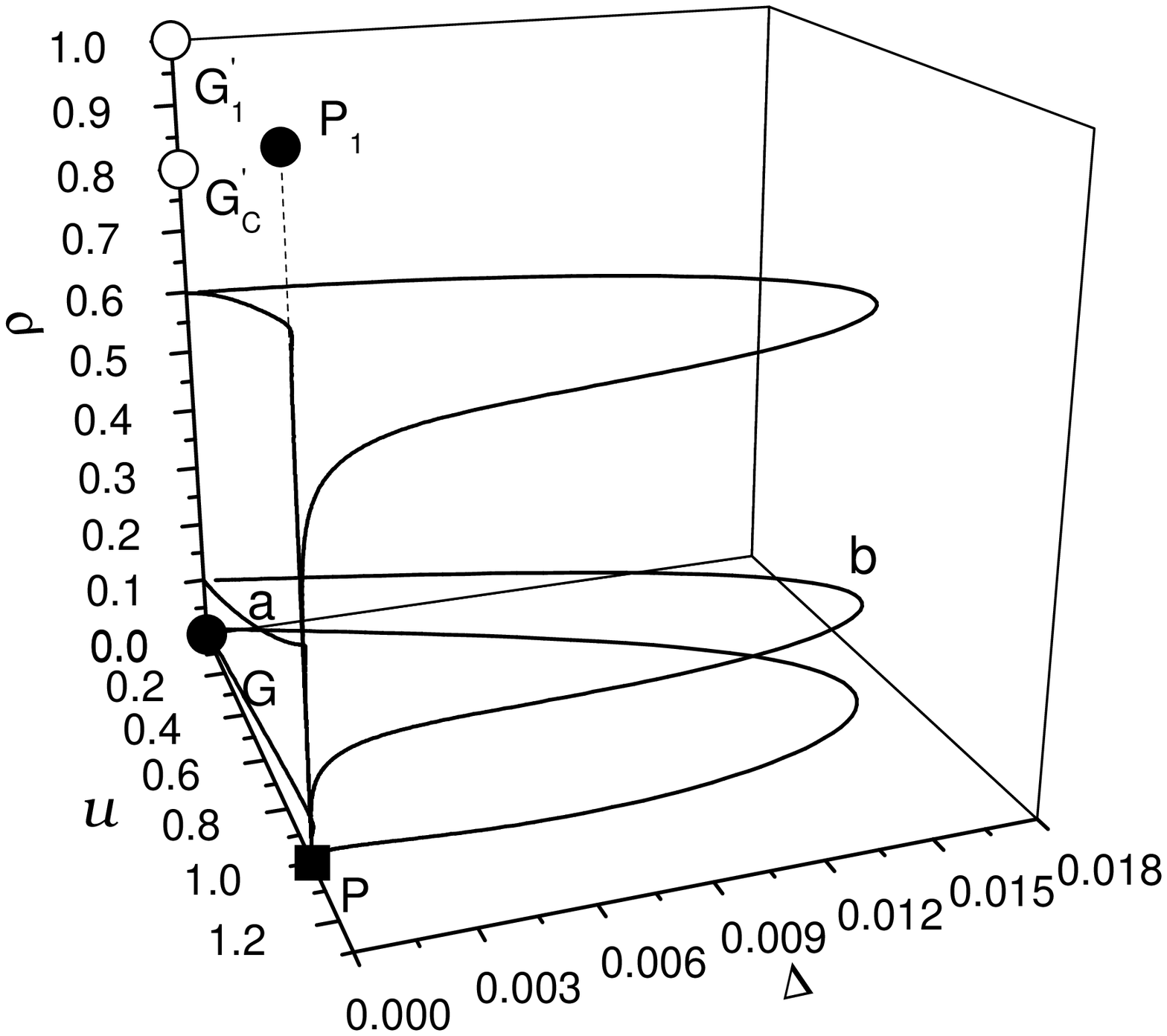}} \end{center}
\caption{\label{dflow2} Static (left picture) and dynamic (right
picture) flows of disordered model C for $n=3$ ($n>n_c$). Filled
circles mean unstable FPs with $\gamma^*=0$, blank circles
indicate projections of unstable FPs with nonzero $\gamma^*$,
while the filled square denotes the stable FP.}
\end{figure}

The second important case we consider here is the Heisenberg
systems $n=3$. In asymptotics their critical behaviour is
characterized by the FP {\bf P}. The static flows and projections
of the dynamic flows in this case are shown in Fig.~\ref{dflow2}.
There exists a considerable difference to the one-loop picture. In
the one-loop approximation the stable FP {\bf P} is marginal: for
this FP every value of $\rho^*$ between 0 and 1 is allowed.
Therefore starting from different initial values flows approach
different $\rho^*$. In contrast, Fig.~\ref{dflow2} shows that
within two-loop approximation all flows remain near the initial
values of $\rho$. Only when the flows reach the region $u=u^*$,
$\Delta=0$,  $\rho$ drops down to its FP value $\rho^*=0$. This
may be attributed to the marginality in one-loop order. The
behaviour of $z_{\rm eff}$ corresponding to flow a of
Fig.~\ref{dflow2} is shown in right picture of Fig.~\ref{z1}. We
present the dependencies of the effective dynamical critical
exponent on the flow parameter only for one set of values
$\gamma(\ell_0)$ and $\rho(\ell_0)$ (but for two different ratios
$u(\ell_0)/\Delta(\ell_0)$ ) because for other values
$\gamma(\ell_0)$ and $\rho(\ell_0)$ we observe similar behaviour:
the peak increases for larger $\gamma(\ell_0)$ and/or
$\rho(\ell_0)$. The experimental investigations give for a
disordered Heisenberg magnet a value of the critical exponent
$z=2.3\pm0.1$ \cite{Alba01}, that is larger than the value of the
dynamical exponent for pure model A with $n=3$ (in our case
$z=2.05$). This might be a consequence of the measurements
performed in the non-asymptotic region.

\section{\label{V} Conclusions}

We have studied the critical dynamics of  model C in the presence
of disorder.

Already one-loop order gives a  qualitatively correct picture for
suitable numbers of OP components $n$. In the asymptotics the
conserved density is decoupled from the OP as it was expected.
However the non-asymptotic critical behaviour of model C is
strongly influenced by the presence of the static coupling between
the OP and the conserved density as well as by the disorder. When
approaching the asymptotic region the dynamical critical exponent
usually shows a maximum starting from the van Hove value $z=2$ in
the background.  For the case $n<n_c$ $(n_c=4)$ the effective
exponent might go after the maximum again through a minimum after
it reaches its asymptotic value. It might be also the case that
one observes the asymptotic value of the dynamical exponent
although the parameters of the system are far from their
asymptotic values. For systems with OP components $n>n_c$ only a
maximum is observed.

These general observations remain  qualitatively the same in
two-loop approximation and are modified  only quantitatively. In
particular, the value of the marginal dimension $n_c$ lies between
1 and $2$ at $d=3$.

The existence of  numerous  FPs leads to a different cross-over
behaviour of RG  flows. Therefore the
 dynamical critical exponent can
assume different values approaching asymptotics.  For the case
$n<n_c$ $z_{\rm eff}$ can either exceed the asymptotic
 value showing a peak or reaches its asymptotic value mechanically. For a system with order
parameter dimension $n>n_c$ $z_{\rm eff}$ always shows a
non-monotonous behaviour.

\section*{Acknowledgements} This work was supported by Fonds zur
F\"orderung der wissenschaftlichen Forschung under Project No.
P16574

\appendix

\section{\label{A} Dynamical functionals and perturbation expansion}

The model defined by expressions (\ref{eq_mov2})-(\ref{r_dist})
within Bausch-Janssen-Wagner formulation \cite{Bausch76} turns out
to be described by an unrenormalized Lagrangian:
\begin{eqnarray}\label{L}
\fl {\mathcal L}=\int d^dx dt \Bigg\{
{-}\mathring{\Gamma}\sum_{i=1}^n{\tilde{\varphi}_{0,i}}{\tilde{\varphi}_{0,i}}{+}
\sum_{i=1}^n{\tilde{\varphi}_{0,i}}
\left({\frac{\partial}{\partial t}}
{+}\mathring{\Gamma}(\mathring{r}{-}
\nabla^2)\right)\varphi_{0,i}+ \mathring{\lambda}_m
\tilde{m}_0\nabla^2\tilde{m}_0\nonumber\\+\tilde{m}_0\left({\frac{\partial}{\partial
t}}-a_m\mathring{\lambda}_m\nabla^2\right)m_0+
\frac{1}{3!}\mathring{\Gamma}\mathring{u}\sum_i\tilde{\varphi}_{0,i}
{\varphi}_{0,i}\sum_j{\varphi}_{0,j} {\varphi}_{0,j}\nonumber\\+
\mathring{\Gamma}V(x)
\sum_i\tilde{\varphi}_{0,i}(x,t){\varphi}_{0,i}(x,t)+
\mathring{\Gamma}\mathring{\gamma}_m
m_0\sum_i\tilde{\varphi}_{0,i} {\varphi}_{0,i}\nonumber\\{-}
\frac{1}{2}\mathring{\lambda}_m\mathring{\gamma}_m
\tilde{m}_0\sum_i\nabla^2{\varphi}_{0,i} {\varphi}_{0,i}\Bigg\}
\end{eqnarray}
 with new auxiliary response fields
${\tilde\varphi}_i(x,t)$. There are two ways to average over the
random potential of disorder $V(x)$ for dynamics. The first way is
the same as in statics and consists in using the replica trick
\cite{Emery75}, where  $N$ replicas of the system are introduced
in order to facilitate configurational averaging of the
corresponding generating functional. Finally the limit $N\to 0$
has to be taken.

However it was established \cite{DeDominicis78} that
renormalization of the replicated Lagrangian leads to the same
results as  the renormalization of an Lagrangian obtained avoiding
the replica trick but taking the mean of the Lagrangian (\ref{L})
with respect to the random potential with distribution
(\ref{r_dist}). The Lagrangian obtained in this way can be written
in the form
 \begin{equation}  \label{Laveraged}
 {\mathcal L}={\mathcal L}_0+{\mathcal L}_{int}   \, ,
\end{equation}
where the Gaussian part ${\mathcal L}_0$ is given by (\ref{L0})
and ${\mathcal L}_{int}$ by  (\ref{Lint}). We perform the
calculations on the basis of the Lagrangian defined by
(\ref{Laveraged}) using the Feynman graph technique.

Response propagators for OP $G(k,\omega)$ and secondary density
$H(k,\omega)$ are equal to
\begin{equation}
\fl
G(k,\omega)=1/(-i\omega+\mathring{\Gamma}(\mathring{\tilde{r}}+k^2))
\qquad\mbox{and}\qquad
H(k,\omega)=1/(-i\omega+a_m\mathring{\lambda}_m k^2)  \,  ,
\end{equation}
 while the correlation propagators $C(k,\omega)$ and $D(k,\omega)$ are equal to
\begin{equation}
\fl
C(k,\omega)=2\mathring{\Gamma}/|-i\omega+\mathring{\Gamma}(\mathring{\tilde{r}}+k^2)|^2
\qquad\mbox{and}\qquad D(k,\omega)=2\mathring{\lambda}_m
k^2/|-i\omega+a_m\mathring{\lambda}_mk^2|^2 \, .
\end{equation}

We proceed within two-loop approximation. In order to obtain the
two-point vertex function
${{\mathring{\Gamma}_{\tilde{\varphi}\varphi}}}^{i,j}(\mathring{\tilde{r}},
\mathring{\tilde
u},\mathring{\Delta},\mathring{\gamma}_m,\mathring{\Gamma},
\mathring{\lambda}_m,a_m,k,\omega)=
{\mathring{\Gamma}_{\tilde{\varphi}\varphi}}(\mathring{\tilde{r}},
\mathring{\tilde
u},\mathring{\Delta},\mathring{\gamma}_m,\mathring{\Gamma},
\mathring{\lambda}_m,a_m,k,\omega)\delta_{i,j}$  one needs to
calculate diagrams of corresponding order. The result of the
calculations  can be expressed in the form:
\begin{eqnarray}\label{form}
\fl \mathring{\Gamma}_{\tilde{\varphi}\varphi}(\xi,k,\omega)=
-i\omega\mathring{\Omega}_{\tilde{\varphi}\varphi}(\xi, k,\omega)+
\mathring{\Gamma}^{st}_{{\varphi}\varphi}(\xi,
k)\mathring{\Gamma}.
 \end{eqnarray}
Here we introduce the correlation length
$\xi(\mathring\tau,\mathring{u},\mathring{\Delta})$, which is
defined by
\begin{eqnarray}
\xi^2=\left.\frac{\partial\ln\mathring{\Gamma}^{st}_{\varphi\varphi}}{\partial
k^2 }\right|_{k^2=0}.
 \end{eqnarray}
The function $\mathring{\Gamma}_{{\varphi}\varphi}$ is the static
two-loop vertex function of a disordered magnet. The structure
(\ref{form}) of the dynamic vertex function of pure model C was
obtained in \cite{Folk02}.

We can express the two-loop dynamical function
$\mathring{\Omega}_{\tilde{\varphi}\varphi}$ in the form:
\begin{eqnarray}\label{om}
\fl \mathring{\Omega}_{\tilde{\varphi}\varphi}(\xi,k,\omega)=1+
\mathring{\Omega}^1_{\tilde{\varphi}\varphi}(\xi,k,\omega)+
\mathring{\Omega}^2_{\tilde{\varphi}\varphi}(\xi,k,\omega),
 \end{eqnarray}
where the one loop contribution has the structure
\begin{eqnarray}\label{om1}
\fl
\mathring{\Omega}^1_{\tilde{\varphi}\varphi}(\xi,k,\omega)=4\mathring{\Delta}\mathring{\Gamma}\int_{k'}
\frac{1}{(-i{\omega}+\mathring{\Gamma}(\xi^{-2}+k'^2))(\xi^{-2}{+}k'^2)}+
{\gamma}\mathring{\Gamma}I_C(\xi,k,\omega),
 \end{eqnarray}
while the two-loop contribution is of the form:
\begin{eqnarray}\label{om2}
\fl \mathring{\Omega}^2_{\tilde{\varphi}\varphi}(\xi,k,\omega)=
\frac{n+2}{18}\mathring\Gamma {\mathring u}^2{\mathring
W}^{(A)}_{{\tilde\varphi}\varphi}(\xi,k,\omega)-\frac{n+2}{3}\mathring\Gamma
{\mathring u}{{\mathring\gamma}^2}{\mathring
C}^{(T3)}_{{\tilde\varphi}\varphi}(\xi,k,\omega)+\mathring\Gamma
{{\mathring\gamma}^4}{\mathring
S}_{{\tilde\varphi}\varphi}(\xi,k,\omega)-\nonumber\\4\frac{n{+}2}{3}\mathring\Gamma
{\mathring u}{\mathring\Delta}{\mathring
W}^{(CD2)}_{{\tilde\varphi}\varphi}(\xi,k,\omega){+}16\mathring\Gamma
{\mathring\Delta}^2\left({\mathring
W}^{(CD3)}_{{\tilde\varphi}\varphi}(\xi,k,\omega){+}{\mathring
W}^{(CD4)}_{{\tilde\varphi}\varphi}(\xi,k,\omega)\right){+}\nonumber\\4\mathring\Gamma
{\mathring\Delta}{\mathring\gamma}^2\left({\mathring
W}^{(CD5)}_{{\tilde\varphi}\varphi}(\xi,k,\omega)+{\mathring
W}^{(CD6)}_{{\tilde\varphi}\varphi}(\xi,k,\omega)+2{\mathring
W}^{(CD7)}_{{\tilde\varphi}\varphi}(\xi,k,\omega)\right),
 \end{eqnarray}
with the rescaled coupling
$\mathring{\gamma}=\mathring{\gamma}_m/{\sqrt {a_m}} $.

Expressions for the integral $I_C$ and for ${\mathring W}^{(A)}$,
${\mathring C}^{(T3)}$ and ${\mathring S}$ of pure model C are
given in Appendix A.1 in Ref.~\cite{Folk04}, the contributions
${\mathring W}^{(CDi)}$ are the following:

\begin{eqnarray}
\fl {\mathring
W}_{{\tilde\varphi}\varphi}^{(CD2)}(\xi,k,\omega)=\int_{k'}\int_{k''}
\frac{1}{(\xi^{-2}+k'^2)(\xi^{-2}+k''^2)(\xi^{-2}+(k+k'+k'')^2) }
\nonumber\\
 \times\frac{1}{(-i\omega+\mathring\Gamma(\xi^{-2}+(k+k'+k'')^2))},
\end{eqnarray}

\begin{equation}
\fl {\mathring
W}_{{\tilde\varphi}\varphi}^{(CD3)}(\xi,k,\omega)=\int_{k'}\int_{k''}
\frac{\mathring\Gamma}{(\xi^{-2}+k''^2)(-i\omega+\mathring\Gamma(\xi^{-2}+k'^2))^2
(-i\omega+\mathring\Gamma(\xi^{-2}+k''^2))},
\end{equation}

\begin{eqnarray}
\fl {\mathring
W}_{{\tilde\varphi}\varphi}^{(CD4)}(\xi,k,\omega)=\int_{k'}\int_{k''}
\frac{1}{(\xi^{-2}+k'^2)(-i\omega+\mathring\Gamma(\xi^{-2}+(k+k'+k'')^2))}
\nonumber\\
\times\Bigg[\frac{1}{(\xi^{-2}+k''^2)}\left(\frac{1}{(\xi^{-2}+(k+k'+k'')^2)}+
\frac{\mathring\Gamma}{-i\omega+\mathring\Gamma(\xi^{-2}+k''^2)}
\right)\nonumber\\+\frac{{\mathring\Gamma}^2}{
(-i\omega+\mathring\Gamma(\xi^{-2}+k'^2))
(-i\omega+\mathring\Gamma(\xi^{-2}+k''^2))}\Bigg],
\end{eqnarray}

\begin{eqnarray}
\fl {\mathring
W}_{{\tilde\varphi}\varphi}^{(CD5)}(\xi,k,\omega){=}\!\int_{k'}\!\!\int_{k''}\!\!
\frac{{\mathring\Gamma}^2}{(\xi^{-2}{+}k''^2)({-}i\omega{+}\mathring\Gamma(\xi^{-2}{+}k'^2))^2
}\nonumber\\ \times
\frac{1}{({-}i\omega{+}\mathring\Gamma(\xi^{-2}{+}k''^2){+}\mathring\lambda(k'{+}k'')^2)},
\end{eqnarray}

\begin{eqnarray}
\fl {\mathring
W}_{{\tilde\varphi}\varphi}^{(CD6)}(\xi,k,\omega){=}\int_{k'}\int_{k''}
\frac{{\mathring\Gamma}^2}{(\xi^{-2}{+}k''^2)
(-i\omega{+}\mathring\Gamma(\xi^{-2}{+}k'^2){+}\mathring\lambda(k{+}k')^2)^2}
\nonumber\\ \times\frac{1}{
(-i\omega{+}\mathring\Gamma(\xi^{-2}{+}k''^2){+}\mathring\lambda(k{+}k')^2)},
\end{eqnarray}

\begin{eqnarray}
\fl {\mathring
W}_{{\tilde\varphi}\varphi}^{(CD7)}(\xi,k,\omega)=\int_{k'}\int_{k''}
\frac{1}{(\xi^{-2}+(k+k'+k'')^2)(-i\omega+\mathring\Gamma(\xi^{-2}+k'^2))}\nonumber\\
\times\frac{1}{
(-i\omega+\mathring\Gamma(\xi^{-2}+k''^2)+\mathring\lambda(k'+k'')^2)}
\nonumber\\ \times\Bigg[\frac{\mathring\Gamma}{(\xi^{-2}+k'^2)}+
\frac{{\mathring\Gamma}^2}
{(-i\omega+\mathring\Gamma(\xi^{-2}+(k+k'+k'')^2)+\mathring\lambda(k'+k'')^2)}\Bigg],
\end{eqnarray}
where we use rescaling $\mathring\lambda=a_m\mathring{\lambda}_m$.

\section{\label{B}Renormalizing factors}
We use the minimal subtraction RG scheme \cite{thoft} for
renormalization. In the definition of the renormalizing factors we
follow  Ref. \cite{Folk04}.

For renormalization of the OP $\vec\varphi$, fourth-order
couplings $u,\,\Delta$ and correlation functions with $\varphi^2$
insertion we introduce the renormalizing factors $Z_{\varphi}$,
$Z_u$, $Z_{\Delta}$ and $Z_{\varphi^2}$ respectively via the
relations:
\begin{equation}\label{ph}
\fl \vec{\varphi}_0=Z^{1/2}_{\varphi}\vec{\varphi}, \quad
{\mathring u}=\mu^{\varepsilon} Z_{\varphi}^{-2}Z_{u}u A_d^{-1},
\quad {\mathring \Delta}=\mu^{\varepsilon/2}
Z_{\varphi}^{-2}Z_{\Delta}\Delta A_d^{-1}, \quad
\frac{1}{2}|\varphi_0|^2= Z_{\varphi^2}\frac{1}{2}|\varphi|^2,
\end{equation}
where $\mu$ is the scale and $\varepsilon=4-d$.

The renormalization of $\mathring r$ is performed
 via the relation
\begin{equation}
{\mathring r}=Z_{\varphi}^{-1}Z_r r.
\end{equation}
The renormalizing factor $Z_r$ is connected to $Z_{\varphi^2}$ by
the relation $Z_{\varphi^2}=Z_{\varphi}^{-1}Z_{r}$ therefore
within minimal subtraction scheme one does not need to consider
renormalization for correlation functions containing $\varphi^2$
insertions explicitly. However a correlation function containing
two insertions $<\varphi^2 \varphi^2>$ needs additive
renormalization $A_{\varphi^2}$.

The renormalization of dynamic quantities is introduced similar to
statics. Renormalizing  factors for the dynamic field
$\vec{\tilde\varphi}$, $Z_{\tilde\varphi}$, and kinetic
coefficient $\Gamma$, $Z_{\Gamma}$, are introduced via
\begin{equation}
\vec{\tilde \varphi}_0{=}Z^{1/2}_{\tilde \varphi}\vec{\tilde
\varphi}, \quad \mathring \Gamma{=}Z_{\Gamma}\Gamma.
\end{equation}

The factor $Z_{\Gamma}$ in the last equation contains a static
contribution $Z_{\varphi}$, which can be separated:
\begin{eqnarray}
Z_{\Gamma}=Z_{\varphi}^{1/2}Z_{\tilde\varphi}^{-1/2}Z^{(d)}_{\Gamma}.
\end{eqnarray}
Since in the dynamic model mode coupling terms are absent
$Z_{\Gamma}^{(d)}=1$. Therefore
\begin{eqnarray}\label{relatZg}
Z_{\Gamma}=Z_{\varphi}^{1/2}Z_{\tilde\varphi}^{-1/2}.
\end{eqnarray}

In model C one needs to introduce additional renormalizing factors
for the secondary density $m$ and its coupling parameter $\gamma$.
They are renormalized  similar to $\varphi$ and $u$ in Eq.
(\ref{ph}):
\begin{eqnarray}
a_{m}^{1/2}m_0=Z_m m,\quad \label{gam}
&a_{m}^{-1/2}{\mathring\gamma}_m
=\mu^{\varepsilon/2}Z_{\varphi}^{-1}Z_m^{-1}Z_{\gamma}\gamma
A_d^{1/2}. \end{eqnarray} There are relations  connecting the
static $Z$-factors of model C to the static $Z$-factors of the
Landau-Ginzburg-Wilson model by integrating out the secondary
field $m$ in the Hamiltonian (\ref{hamilt1}). Thus the
renormalizing factor of $\gamma$ is  determined by:
\begin{equation}  \label{zfaktorgamma}
Z_{\gamma}=Z^2_{m}Z_{\varphi}Z_{\varphi^2}.
\end{equation}
Therefore Eq. (\ref{gam}) can be rewritten as
\begin{eqnarray}
&a_{m}^{-1/2}{\mathring\gamma}_m=\mu^{\varepsilon/2}Z_{\varphi^2}Z_m\gamma
A_d^{-1/2}.& \end{eqnarray} The additive renormalization
$A_{\varphi^2}$ of the specific heat of the Landau-Ginzburg-Wilson
model determines the renormalizing factor $Z_m$ via the relation
\begin{eqnarray}  \label{zfaktorm}
&Z_{m}^{-2}(u,\Delta,\gamma)=1+\gamma^2 A_{\varphi^2}(u)&.
\end{eqnarray}

Since the secondary density is conserved, no new renormalizing
factor is needed for the dynamic auxiliary density $\tilde m$. It
renormalizes by:
\begin{eqnarray}
&a_{m}^{-1/2}{\tilde m}_0=Z_m^{-1}\tilde m&.
 \end{eqnarray}
The kinetic coefficient $\lambda$ renormalizes as
\begin{eqnarray}
a_{m}{\mathring \lambda}_m=Z_{\lambda}\lambda.
\end{eqnarray}
Similar to  $Z_{\Gamma}$ its static contribution can be separated:
\begin{equation}
 Z_{\lambda}=Z^2_{m}Z_{\lambda}^{(d)} .
 \end{equation}
Since no mode coupling terms are present in  model C,
$Z_{\lambda}^{(d)}=1$, therefore
\begin{eqnarray}\label{relatZl} Z_{\lambda}=Z^2_{m} .
 \end{eqnarray}

 Renormalizing $\mathring{\Gamma}_{\tilde\varphi\varphi}$ we obtain
the two-loop renormalizing factor $Z_{\tilde \varphi}$ in the
form:
\begin{eqnarray}\label{zfac}
\fl Z_{\tilde
\varphi}=1{-}8\frac{\Delta}{\varepsilon}-2\frac{\gamma^2}{\varepsilon}\frac{w}{1+w}
+\frac{1}{\varepsilon^2}\Big[\left({\gamma^2}\frac{w}{1+w}\left(\frac{1}{w+1}
-\left(\frac{n}{2}-1\right)\right)-\frac{n+2}{3}u\right){\gamma^2}\frac{w}{1+w}
\nonumber\\ +
4\Delta{\gamma^2}\!\left(\!\frac{w}{1{+}w}\!\right)^2\!\!{+}20\Delta{\gamma^2}\frac{w}{1{+}w}\!+96\Delta^2
{-}4\frac{n{+}2}{3}u\Delta\Big]{+}
\frac{1}{2\varepsilon}\Bigg\{\!\Bigg[\frac{n{+}2}{3}u\left(\!\!1{-}3\ln\frac{4}{3}\!\right)
\nonumber\\ {+} {\gamma^2}\!\frac{w}{1{+}w}\Bigg(\!\frac{n}{2}{-}
\frac{w}{1{+}w}{-}\frac{3(n{+}2)}{2}\ln\frac{4}{3}{-}
\frac{1{+}2w}{1{+}w}\ln\frac{(1{+}w)^2}{1{+}2w}\!\Bigg)\Bigg]{\gamma^2}\frac{w}{1{+}w}{+}
3\frac{n{+}2}{3}u\Delta\nonumber\\{-}44\Delta^2{-}12\Delta{\gamma^2}\frac{w}{1{+}w}\Bigg\}{-}
\frac{n{+}2}{12}\frac{u}{\varepsilon}\left(\ln\frac{4}{3}{-}\frac{1}{12}\right){+}
\nonumber\\
4\frac{\Delta}{\varepsilon}{\gamma^2}\frac{w}{1+w}\left[\frac{w}{2}\ln\frac{w}{1+w}-
 \frac{3}{2}\ln(1+w) -\frac{1}{2}\frac{w}{1+w}\ln w\right]
 \end{eqnarray}


\section*{References}
\begin{thebibliography}{99}

\bibitem{review}
For  recent reviews see e.g.:  R.~Folk, Yu.~Holovatch,
T.~Yavors'kii, Physics - Uspekhi {\bf 46} 169 (2003) [Uspekhi
Fizicheskikh Nauk {\bf 173} 175],  preprint {\it
cond-mat/0106468}; A. Pelissetto, E. Vicari,  Phys. Rep. {\bf 368}
549 (2002), preprint {\it cond-mat/0012164}

\bibitem{Harris74} A. B. Harris, J. Phys. C: Solid State Phys. {\bf 7}, 1671 (1974)

\bibitem{note1} This is the way in which the Harris criterion was originally formulated.
However afterwards it has been often interpreted as a  prediction
of a change in asymptotic criticality if $\alpha_{pure}>0$. We
thank W. Janke for attracting our attention to the original
formulation and later sloppy interpretations of the Harris
criterion. For the cases treated here when $\alpha_{pure}>0$ a new
universality class is found.


\bibitem{Bervillier86}
C. Bervillier, Phys. Rev. B {\bf 34} 8141 (1986); M. Dudka, Yu.
Holovatch, T. Yavors'kii, J. Phys. Stud.
 {\bf 5} 233 (2001).

\bibitem{Halperin74}
B. I. Halperin, P. C. Hohenberg, and S.-k. Ma, Phys. Rev. B {\bf
10}, 139 (1974).

\bibitem{Krey76}
U. Krey, Phys. Lett. {\bf 57A} 215 (1976);
\bibitem{Krey77}
U. Krey, Z. Physik B {\bf26}, 355 (1977).

\bibitem{Lawrie84} I. D. Lawrie, and V. V. Prudnikov, J. Phys. C {\bf 17},
1655 (1984).

\bibitem{FoMo03}  R. Folk and G. Moser,
Phys. Rev. Lett. 91, 030601 (2003)

\bibitem{Folk04}
R. Folk, G. Moser, Phys. Rev. E {\bf 69}, 036101 (2004).


\bibitem{Grinstein77}
G.Grinstein, S.-k. Ma, and G. Mazenko, Phys. Rev. B {\bf 15} 258
(1977).

\bibitem{Prudnikov92}
V. V. Prudnikov, A. N. Vakilov, Sov. Phys. JETP  {\bf 74}, 990
(1992) [Zh. Eksp. Teor. Fiz. {\bf 101}, 1853 (1992)].

\bibitem{Oerding95}
K. Oerding and H. K. Janssen, J. Phys. A: Math. Gen. {\bf 28},
4271 (1995).

\bibitem{Janssen95}
H. K. Janssen, K. Oerding, and E. Sengenspeick, J. Phys. A: Math.
Gen. {\bf 28}, 6073 (1995).

\bibitem{Dudka03}
M. Dudka, R. Folk, Yu. Holovatch, and D. Ivaneiko, J. Magn. Magn.
Mater. {\bf 53} 243 (2003).

\bibitem{Perumal03}
A. Perumal, V. Srinivas, V. V. Rao, R. A. Dunlap, Phys. Rev. Lett.
{\bf 91} 137202 (2003).

\bibitem{Berche04} P. E. Berche, C.Chatelain, B. Berche, W. Janke, Eur. Phys. J. B,
{\bf 39}, 463 (2004)

\bibitem{Calabrese04}
P. Calabrese, P. Parruccini, A. Pelissetto, E. Vicari, Phys. Rev.
E {\bf 69} 036120 (2004).

\bibitem{Berche05} B. Berche, P. E. Berche, C.Chatelain,  W. Janke, Condens.
Matter Phys. {\bf 8}, 47 (2005)

\bibitem{sqrteps} D. E. Khmel'nitskii,  Zh. Eksp. Teor. Fiz.  { 68}  (1975) 1960 [
Sov. Phys. JETP  { 41} (1975) 981]; A.~B. Harris, T.~C.~Lubensky,
{Phys. Rev. Lett.}  { 33}  (1974) 1540; T. C. Lubensky, Phys. Rev.
B { 11}  (1975) 3573.

\bibitem{Grinstein78} G. Grinstein,  A. Luther, Phys. Rev.~B {\bf 13}  (1976) 1329.

\bibitem{Halperin77}
B. I. Halperin, P. C. Hohenberg,  Rev. Mod. Phys. {\bf 49}, 436
(1977).

\bibitem{Emery75}
V. J. Emery, Phys. Rev. B {\bf 11}, 239 (1975).

\bibitem{Bausch76}
R. Bausch, H. K. Janssen, and H. Wagner, Z. Phys. B {\bf 24}, 113
(1976).

\bibitem{DeDominicis78}
C. De Dominicis, Phys. Rev. B {\bf 18}, 4913 (1978).

\bibitem{Folk00}
R. Folk, Yu. Holovatch, T. Yavors'kii,  Phys. Rev. B {\bf 61}
15114 (2000); J. M. Carmona, A. Pelissetto, and E. Vicari Phys.
Rev. B {\bf 61}, 15136 (2000).

\bibitem{moda1lop}
U.Krey, Z. Phys. B {\bf 26}, 355 (1977).


\bibitem{Folk02}
R. Folk, G. Moser, Acta Physica Slovaca {\bf 52}, 285 (2002).

\bibitem{thoft}   G. 't Hooft,  M. Veltman, { Nucl. Phys. B}  {\bf 44} (1972) 189;
G.~'t Hooft, { Nucl. Phys. B}  {\bf 61}  (1973) 455.







\bibitem{latter}
M. Dudka, R. Folk, Yu. Holovatch, G. Moser, Phys. Rev. E {\bf 72}
036107 (2005).





\bibitem{Kyriakidis96}
J. Kyriakidis and D. J. W. Geldart {\bf 53}, 11572 (1996).




\bibitem{Wilson72} K. Wilson, M. E. Fisher, Phys. Rev.
Lett. { 28} (1972)  240.


\bibitem{Schloms87}
R. Schloms and V. Dohm, Europhys. Lett. { 3} (1987) 413; R.
Schloms and V. Dohm, Nucl. Phys. B {328} (1989)  639.



\bibitem{Baker78}
G. A. Baker, Jr., B. G. Nickel, D. I. Meiron, Phys. Rev.~B { 17},
 (1978) 1365.


\bibitem{Watson74}
P. J. S. Watson,  J. Phys. A  { 7} (1974) L167.

\bibitem{Baker81}
G. A. Baker, Jr., P. Graves-Morris, {\em Pad\'e Approximants}
(Addison-Wesley: Reading, MA, 1981)




 \bibitem{Rosov92}
N. Rosov, C. Hohenemser, M. Eibsch\"utz,  Phys. Rev. B {\bf 46},
3452 (1992).

\bibitem{note2}
Compare with their analytic form that can be obtained with the
help of Eq. (\protect\ref{zpt}).

\bibitem{Alba01}
M. Alba, S. Pouget, J. Magn. Magn. Mater {\bf 226-330} 542 (2001).

\end{thebibliography}
\end{document}